\documentclass[fleqn,usenatbib]{mnras}

\usepackage{newtxtext,newtxmath}
\usepackage[T1]{fontenc}
\usepackage{ae,aecompl}
\usepackage{graphicx}
\usepackage{amsmath}
\usepackage{amssymb}

\usepackage[usenames,dvipsnames,table,xcdraw]{xcolor}
\usepackage{wasysym}
\usepackage{widetext}
\usepackage{float}
\usepackage{subcaption}
\usepackage{array}
\usepackage{multirow}
\usepackage{amstext}

\newcommand{\Romer}{R{\o}mer }
\newcommand{\order}{\mathcal{O}}
\newcommand{\tempo}{{\texttt{TEMPO2}}}
\newcommand{\degree}{^{\circ}}
\newcommand{\BesselJ}{\text{J}}
\providecommand{\tabularnewline}{\\}

\newcommand{\good}[1]{#1}	

\title[Periastron Advance in Small-$e$ Binary Pulsars]{Exploring the Effect of Periastron Advance in Small-Eccentricity Binary Pulsars}

\author[A. Susobhanan et al.]{
Abhimanyu Susobhanan,$^{1}$\thanks{E-mail: s.abhimanyu@tifr.res.in}
Achamveedu Gopakumar,$^{1}$
Bhal Chandra Joshi$^{2}$
and
\newauthor{ Ranjan Kumar$^{3}$}   \\
$^{1}$Department of Astronomy and Astrophysics, Tata Institute of Fundamental Research, Mumbai 400005, India\\
$^{2}$National Centre for Radio Astrophysics (Tata Institute of Fundamental Research), Pune 411007, India\\
$^{3}$Department of Physics and Astronomy, National Institute of Technology, Rourkela 769008, India}

\date{Accepted XXX. Received YYY; in original form ZZZ}

\pubyear{2018}

\begin{document}

\label{firstpage}
\pagerange{\pageref{firstpage}--\pageref{lastpage}}
\maketitle

\begin{abstract}
Short-orbital period small-eccentricity binary pulsars can, in principle, experience substantial advance of periastron.
We explore the possibility of measuring this effect by implementing a timing model, {\ttfamily ELL1k},
in the popular {\ttfamily TEMPO2} pulsar timing package.
True secular variations in the Laplace-Lagrange parameters, present in our {\ttfamily ELL1k} model, 
can lead to measurable timing residuals while pursuing decade-long timing campaigns using the existing
{\ttfamily ELL1} timing model 
\good{of \citet{Lange2001}},
especially for binaries exhibiting significant periastron advance. 
We also list 
{the} 
\good{main} differences between our approach 
\good{and various implementations of the \texttt{ELL1} model present in \good{both} \texttt{TEMPO} and \texttt{TEMPO2}}
packages. 
Detailed {\ttfamily TEMPO2} simulations suggest the possibility of 
constraining the apsidal motion constant 
of pulsar companions in certain observed binary pulsars  
with minuscule eccentricities such as PSR J1719$-$1438.
Fortunately, the {\ttfamily ELL1k} timing model does not pose any challenges to 
the on-going Pulsar Timing Array campaigns that routinely  employ the {\ttfamily ELL1} timing model. 
\end{abstract}

\begin{keywords}
pulsars: general -- binaries: general  
-- gravitation -- relativity
\end{keywords}



\section{Introduction}
\label{sec:Intro}
Presently, binary pulsars provide the most accurate laboratories to test general relativity (GR) in quasi-stationary strong field regime \citep{Wex2014}.
This is mainly because of the technique of pulsar timing which requires 
an accurate prescription for determining the pulse phase as a function of time.
In the case of binary pulsars like PSR B1913+16, this technique essentially provided an ideal clock for probing the nature of  relativistic gravity \citep{Taylor1979}.
Pulsar timing demonstrated the decay of orbital period in neutron star binaries 
which provided the first observational evidence for the existence of gravitational 
waves \citep{Tayor_Nobel_lecture}.
Additionally, the on-going Pulsar Timing Array (PTA) experiments aim to detect low-frequency
gravitational waves in the $10^{-9} \,-\, 10^{-10}$ Hz frequency range \citep{vlh+16}.
These efforts require accurate timing of millisecond pulsars (MSPs) due to their 
exquisite rotational stability.

Roughly, 10\% of the  
\good{over 2600 currently known} pulsars exist  
in binary systems with companions in various stages of stellar evolution \citep{Manchester1993_ATNF_Catalog}.
Accurate timing of pulsars in such systems requires
prescriptions to model delays in the times of arrival (TOAs) of pulses
in an inertial barycentric frame 
due to the orbital motion of the pulsar and its companion \citep{Blandford1976,Edwards2006_TEMPO2}.
For double neutron star binaries in eccentric orbits, a number of 
relativistic orbital effects contribute to such delays \citep{DamourDeruelle1986}.
This ensures that the binary orbit is specified by a number of post-Keplerian 
parameters in addition to the regular  Keplerian parameters, namely,
{the orbital period} $P_a$, 
{argument of periastron} $\omega$, 
{eccentricity} $e$, 
{time of periastron passage} $T_0$ and 
{the projected semi-major axis} \good{$x$}.
Timing of binary pulsars allow us to probe quasi-stationary strong gravitational fields 
due to our ability to measure their post-Keplerian parameters.
However, timing of MSPs in binaries requires special 
care as a good fraction of them are in near-circular orbits. 

For pulsar binaries with tiny orbital eccentricities, the TOAs do not 
prominently depend on Keplerian parameters $\omega$ and $T_0$.
This results in large uncertainties in the usual $\chi^2$ estimation of 
 $\omega$ and $T_0$.
This prompted Norbert Wex to describe such orbits in terms of the first 
 and second Laplace-Lagrange (LL) parameters and certain time of ascending node passage \good{$T_{\ascnode}$} \citep{Lange2001,Edwards2006_TEMPO2}.
These parameters, namely 
$\epsilon_1 = e\, \sin \omega$, 
$\epsilon_2 = e\, \cos \omega$ and 
$T_{\ascnode} = T_{0} - \frac{\omega}{n}$, where $n$ is the mean motion, 
replace the regular Keplerian parameters $e$, $\omega$ and $T_0$.
Clearly, the use of rectangular components of 
the eccentricity vector, 
given by $e\,\cos \omega$ and $e\,\sin \omega$, to represent the periastron of the elliptical orbit is influenced by 
its use in celestial mechanics \citep{Pannekoek1948}.

  The {\ttfamily ELL1} timing model, detailed in \citet{Lange2001}, incorporates the effects of \Romer and 
 Shapiro delays while neglecting the Einstein delay contributions.
 Additionally, {linear-in-time} variations of the sidereal orbital period, 
 Laplace-Lagrange parameters and projected semi-major axis were introduced 
 to model secular changes in these parameters.
 In the {\tempo} implementation of this timing model, one employs 
 ${\epsilon_{10}}$, $\epsilon_{20}$, $\dot \epsilon_1$, $\dot \epsilon_2$ and 
 $T_{\ascnode}$ 
  as fitting parameters in the place of 
 $e_0$, $\omega_0$, $\dot e$, $\dot \omega$ and $T_0$
  employed 
 in the \texttt{DD} timing model for eccentric binaries \citep{DamourDeruelle1986,Edwards2006_TEMPO2}.
 This is what one gathers from a detailed study of the appendix in \citet{Lange2001} and its {\tempo} implementation \citep{Edwards2006_TEMPO2}.
It should be noted that the {\ttfamily ELL1} model incorporates only the first order terms in orbital eccentricity.
Very recently, it was pointed out by \citet{Zhu2018} that higher order eccentricity contributions to the \Romer delay should be 
 relevant while timing 
nearly-circular wide-orbit binary pulsars.
\good{These corrections were implemented in the \texttt{ELL1+} model, an extension of \texttt{ELL1}.}
However, TOA measurement errors are expected to be much larger than the higher order $e$ corrections to the \Romer delay
for short orbital period binaries: systems of our current interest.
 
  In this paper, we explore the implications of 
  restricting the temporal variations of  LL parameters to be linear-in-time, pursued 
  in \citet{Lange2001} and implemented in {\tempo} \citep{Edwards2006_TEMPO2}.
   It turns out that 
  the time evolutions of rectangular components of the
  orbital eccentricity vector are more general.
   This is because the rate of periastron advance per 
  orbit need not be negligible for compact pulsar binaries with tiny 
  orbital eccentricities as evident from equations (1-3) in \citet{Willems2008}.
    The more general time evolutions for the LL parameters can have 
    observational implications due to the following reasons.
  These days it is indeed possible to 
  measure TOAs with 
  \good{$\sim$}100 ns uncertainties due to the 
 advent of real-time coherent de-dispersion 
 \citep{Hankins1987}
 and availability of large collecting area telescopes like the Giant Metre-wave Radio Telescope equipped with 400 MHz wide-band receivers \citep{Reddy2017}.
 The TOA precision is likely to improve further with the advent of telescopes like
 the Five hundred meter Aperture Spherical Telescope (FAST; \citet{Nan2011}) 
  and the Square Kilometre Array (SKA; \citet{Combes2015}). 
  Secondly, the secular systematics in timing residuals are expected to increase with observation span 
  for timing campaigns lasting decades, such as those employed in various PTA  experiments \citep{vlh+16}.
This is relevant for us as many PTA millisecond pulsars
are in small-eccentricity binary orbits and are timed by  
the {\ttfamily ELL1} timing model. 
The final consideration is based on the possibility that the present {investigation}
may allow us to extract additional astrophysical information from 
certain binary pulsar systems like PSR J1719$-$1438 \citep{Bailes2011}.
This is because the classical contributions to the periastron advance 
may dominate over its general relativistic counterpart in a number of 
recently discovered MSP binaries.
If this is indeed the case, the long-term timing of such tiny-eccentricity pulsar binaries may allow us to determine the apsidal 
motion constant of pulsar companions in such systems.

  After completion of our numerical {\tempo} experiments, we came to know that Norbert Wex had implemented a similar 
  extension to his {\ttfamily ELL1} timing model in {\texttt{TEMPO}} \citep{tempo1_ref}.
Unfortunately, no public documentation exists for this implementation.
A close inspection reveals that 
our timing model and Wex's {\texttt{TEMPO}} implementation\footnote{Available at \url{https://sourceforge.net/p/tempo/tempo/ci/master/tree/src/bnryell1.f}} that extends  his 
{\ttfamily ELL1} timing model, detailed in \citet{Lange2001},  differs by a term.
Our {\tempo} simulations reveal that this additional term is crucial for small-eccentricity binary pulsars 
experiencing substantial periastron advance and therefore should not be ignored.

In 
\good{Section \ref{sec:RomerDelay_ELL1}}, we briefly summarize the {\ttfamily ELL1} timing model.  
How we incorporate `exact' temporal evolutions of the LL parameters 
and why long-term monitoring will be required to distinguish such evolutions from the 
present {\tempo} evolutions for $\epsilon_1$ and $\epsilon_2$ are explained in 
Section \ref{sec:Advance-of-periapsis_ELL1}.
Why PSR J1719$-$1438 should be the most promising source to observe
the effects of periastron advance is explained in 
Section \ref{sec:measurability}. 
This section also details 
how we adapted the {\tempo} software\footnote{Available at \url{https://bitbucket.org/psrsoft/tempo2}} \citep{Hobbs2006_TEMPO2,Edwards2006_TEMPO2} to
explore observational consequences of 
the general time evolutions for $\epsilon_1$ and $\epsilon_2$.
In Section \ref{sec:discussion}, we summarize our results and discuss their implications.
Expressions required to incorporate the {\ttfamily ELL1k} model in {\tempo} are given in Appendix \ref{sec:t2_inputs} 
\good{and a brief comparison of different timing models for near-circular binaries is given in Appendix \ref{sec:ell1_comparison_table}.}

\section{The {\ttfamily ELL1} timing model of Norbert Wex } 
\label{sec:RomerDelay_ELL1}

 In the {\ttfamily ELL1} timing model, the barycentric arrival time of a pulse, $t_{\text{obs}}$,
 is related to its time of emission, $t_{\text{em}}$, by
 \begin{equation}
 t_{\text{obs}} 
 	= t_{\text{em}}  + \Delta_{R}(t_{\text{em}}) \,,
 \end{equation}
where $ \Delta_{R}(t_{\text{em}})$ is the \Romer delay  due to the orbital motion of the pulsar 
around its companion.
The standard expression for the \Romer delay associated 
with an eccentric Keplerian orbit  is given by \citep{Blandford1976}
\begin{equation}
\label{eq:RomerDelay}
\Delta_{R} = \frac{x}{c}\left[\left(\cos u - e\right)\sin \omega + \sqrt{1-e^{2}}\sin u \, \cos \omega \right]\,,
\end{equation}
where $x=a_p \sin\iota$ is the semi-major axis of the pulsar orbit projected on to the line of sight and 
$\iota$ is the orbital inclination. 
The Keplerian 
\good{parameters} are 
the eccentric anomaly $u$, 
orbital eccentricity $e$, and the argument of periastron $\omega$. 
\good{The semi-major axis of the pulsar orbit}, 
$a_p$, is related to the semi-major axis of the relative orbit $a$ by the standard relation 
$a_p = \frac{m_c}{M} a$
where the total mass of the binary  $M=m_p+m_c$ is given in terms of 
 the pulsar mass $m_p$ and the companion mass $m_c$.
 The relativistic version of  equation (\ref{eq:RomerDelay}) may be 
 found in \cite{DamourDeruelle1986}.

 It is possible to obtain an explicit analytic expression for the temporal 
 evolution of $ \Delta_{R}$ in the small eccentricity limit.
This is due to the following standard expressions for $\sin u$ and $\cos u$ 
in terms of the mean anomaly $l$
\citep[Ch. 2]{Brouwer1961Ch2} 
\begin{align}
\cos u & = \frac{-e}{2}+2\sum_{s=1}^{\infty}\frac{1}{s}\BesselJ_{s}^{'}(se)\cos sl \nonumber \\
 		& = \cos l + \frac{e}{2}(\cos 2l-1)+\order(e^2)\,, \label{eq:cosu} \\
\sin u & = \frac{2}{e}\sum_{s=1}^{\infty}\frac{1}{s}\BesselJ_{s}(se)\sin sl \nonumber \\
 		& = \sin l + \frac{e}{2} \sin 2l + \order(e^2) \label{eq:sinu} \,,
\end{align}
where $l$ is defined in terms of the mean motion $n$ and the epoch of periastron passage $T_0$ as 
$l=n \left(t-T_{0}\right)$ while $ \BesselJ_{s}(se) $ and $\BesselJ_{s}^{'}(se)$ stand for the Bessel function of the first kind 
and its derivative.
General relativistic corrections to the above expressions, accurate to third post-Newtonian order, are available in \cite{Boetzel2017}.
The mean motion $n$ is related to the orbital (anomalistic) period by 
$n ={2\pi}/{P_{a}}$.
Wex introduced a certain phase $\Phi=l+\omega$  such that equation (\ref{eq:RomerDelay}) 
with the help of equations (\ref{eq:cosu}) and (\ref{eq:sinu}) becomes 
\citep{Lange2001}
\begin{align}
\Delta_{R} = \frac{x}{c} \left[\sin\Phi+\frac{e}{2}\left(\sin2\Phi\:\cos\omega-\left(\cos2\Phi+3\right)\sin\omega\right)\right]
\label{eq:Romer_delay_intermediate}
\end{align}
The above expression, which is accurate to linear order in $e$, takes a simpler form in terms of the Laplace-Lagrange parameters  
$\epsilon_{1} =e\sin\omega$ and 
$\epsilon_{2} =e\cos\omega$, and it reads 
\begin{align}
\label{eq:Romer_Delay_ELL1}
\Delta_{R}=\frac{x}{c}\left[\sin\Phi+\frac{1}{2}\biggl (\epsilon_{2}\sin2\Phi-\epsilon_{1}\left(\cos2\Phi+3\right)\biggr )\right].
\end{align}
A comparison with equation (A6) in \cite{Lange2001} reveals the presence of an additional term 
$ \frac{-3}{2} \frac{x}{c} \epsilon_1$ which was neglected in the {\ttfamily ELL1} timing model.
We observe that this term is absent in both {\tempo} and {\texttt{TEMPO}} implementations
of Wex's timing model for small-eccentricity binary pulsars.
To obtain explicit temporal evolution for $\Phi$ in terms of well-defined parameters, Wex introduced certain 
{\it time of ascending node } $T_{\ascnode} $ to be
\begin{equation}
T_{\ascnode}=T_{0}-\frac{\omega}{n}\,,
\end{equation}
such that the time evolution for the phase is given by  
\good{\begin{align}
\Phi & =l+\omega 
       =n\left( t-T_{0}+\frac{\omega}{n} \right)  = n \left ( t- T_{\ascnode} \right ) \,.
\end{align}}
The introduction of  $T_{\ascnode} $  ensures that the usual 
Keplerian parameters, $e$, $\omega$ and $T_0$ are replaced by parameters
$\epsilon_1$, $\epsilon_2$ and $T_{\ascnode}$, which are more appropriate 
for binaries having tiny orbital eccentricities.

\good{When advance of periastron is present,} it is convenient to introduce certain sidereal angular frequency $n_b = n + \dot \omega$  such that the secular evolution of the phase becomes 
\begin{align}
\Phi & =n_{b}\left(t-T_{\ascnode}\right)+\frac{1}{2}\dot{n}_{b}\left(t-T_{\ascnode}\right)^{2} \,,
\end{align}
where \good{we have also incorporated the orbital frequency derivative} $\dot n_b = \dot n$ due to the dissipative evolution of the orbit.
The secular variations to $x$, $\epsilon_1$ and $\epsilon_2$ are provided 
in the {\tempo} implementation of \texttt{ELL1}
by the following relations 
\citep{Lange2001} 
\begin{align}
x &= x_0 + \dot x \,\left(t-T_{\ascnode}\right) \,, \nonumber \\
\epsilon_1 &= \epsilon_{10} + \dot { \epsilon_1} \,\left(t-T_{\ascnode}\right) \,,  \nonumber \\
\epsilon_2 &= \epsilon_{20} + \dot { \epsilon_2} \,\left(t-T_{\ascnode}\right)\,,
\end{align}
where \good{$T_{\ascnode}$ is now defined as} $T_{\ascnode} = T_0 - \omega_0/n_b $, \good{and $\omega_0$, $x_0$, $\epsilon_{10}$ and $\epsilon_{20}$ are the values of the corresponding parameters at $t=T_{\ascnode}$}.

This ensures that {\ttfamily ELL1} model {in \texttt{TEMPO2}} employs 
$\epsilon_{10}$, 
$\epsilon_{20}$, 
$\dot \epsilon_1$, 
$\dot \epsilon_2$ and
$T_{\ascnode}$ 
as fittable parameters in addition to $x_0$, and $\dot x$ if required.
We note, in passing, that  Einstein delay is not relevant for these systems while the Shapiro delay expression of \citet{Lange2001}
is not altered by our considerations.

 We would like to state explicitly that a prescription for the general temporal evolutions of 
  the LL parameters  in terms of $\dot e$ and $\dot\omega$ is available in  the \texttt{TEMPO} software.
  Unfortunately, this prescription is missing in the currently popular {\tempo} software, and as noted earlier 
  no public documentation exists for the \texttt{TEMPO} implementation that generalizes 
 the above-detailed  \texttt{ELL1} timing model.
In what follows, we detail  how we independently generalized  linear-in-time evolutions 
for $\epsilon_1$ and $\epsilon_2$ while including the above mentioned additional term in the 
expression for the  \Romer delay.

\section{Generalizing the linear-in-time  evolutions for $\epsilon_1$ and $\epsilon_2$ }
\label{sec:Advance-of-periapsis_ELL1}
In this section, 
we describe a simple prescription to model the `exact' time  evolutions for $\epsilon_1$ and $\epsilon_2$ 
due to linear-in-time evolutions for $e$ and $\omega$ with the aim of extracting 
the effect of $\dot \omega $ from binary pulsar timing.
It turns out that  the effect of periastron advance is implicitly present in the {{\ttfamily ELL1}} timing model 
 and this forces the conservative phase evolution  to be $\Phi = n_b \left ( t - T_{\ascnode} \right )$ where 
 $ n_b = n + \dot \omega $. However, it is fairly difficult to constrain $\dot \omega$ from timing observations.
 This is because the times of ascending node passages are provided in terms of $P_b = 2\, \pi/n_b$
 such that the epoch of $N^{\text{th}}$  ascending node  passage reads 
\begin{equation}
T_{\ascnode}^{(N)}=T_{\ascnode}^{(0)} + N\,P_{b} + \frac{1}{2}N^{2}P_{b}\dot{P}_{b}\,,
\end{equation}
where we imposed linear-in-time evolution for $n_b$.
Therefore, one may extract the time derivative of $P_b$ by measuring 
$T_{\ascnode}$ values at widely separated epochs as done, for example, in the case of 
the accreting millisecond X-ray pulsar 
SAX J1808.4$-$3658 \citep{Patruno2012}.
In terms of $P_b$, time evolution of 
$\Phi$ is given by
\begin{align}
\Phi & =2\pi\left[\frac{\tau}{P_{b}}-\frac{1}{2}\dot{P}_{b}\left(\frac{\tau}{P_{b}}\right)^{2}\right]\,,
\end{align}
where we have defined $\tau=t-T_{\ascnode}$ and  neglected higher derivatives of $P_{b}$. 

It is straightforward to show that the general temporal evolutions for the Laplace-Lagrange parameters are given by
\begin{align}
\label{eq:LL_evolution_0}
\epsilon_{1} (t) &= (1+\xi\tau)\left(\epsilon_{10}\cos\dot{\omega}\tau+\epsilon_{20}\sin\dot{\omega}\tau\right),   \nonumber \\
\epsilon_{2} (t) &= (1+\xi\tau)\left(\epsilon_{20}\cos\dot{\omega}\tau-\epsilon_{10}\sin\dot{\omega}\tau\right),
\end{align}
where {$\xi=\dot{e}/e_0$ and} the subscript $0$ stands for parameter values at $\tau=0$ such that 
\begin{align}
\epsilon_{10}& = e_0 \sin \omega_{0} \,,\nonumber \\
\epsilon_{20}& = e_0 \cos \omega_{0} \,.
\end{align}

Note that we let $\omega$ and $e$ to vary linear-in-time, namely $\omega = \omega_0 + \dot \omega \, \tau$ and $e=e_0 + \dot{e} \tau$,
to obtain the above time evolutions for $\epsilon_1$ and $\epsilon_2$.
Further, close inspection of  $d e/dt$  expressions due to 
dominant order gravitational radiation reaction effects and tidal dissipation
allows us to impose $\xi =0$ for these binaries  \citep{Peters1964,Zahn1978}.
For example,  $\dot e$ contributions, due 
to gravitational wave emission,
appear at $(v/c)^5$ order and  are small compared to the periastron advance contribution which occurs at $(v/c)^2$ order where 
$v$ is the orbital velocity. 
This ensures that the timescale $\xi^{-1}$ associated 
with $\dot e$ will be substantially higher than the timescale relevant for the periastron advance. 
However, $\dot e$ contributions should not be neglected 
while testing strong equivalence principle
using  small-eccentricity long-orbital period binary pulsars 
\citep{DamourSchafer1991}.
Neglecting $\xi$, equations (\ref{eq:LL_evolution_0}) become
\begin{align}
\label{eq:LL_evolution}
\epsilon_{1} (t) &= \epsilon_{10}\cos\dot{\omega}\tau+\epsilon_{20}\sin\dot{\omega}\tau,   \nonumber \\
\epsilon_{2} (t) &= \epsilon_{20}\cos\dot{\omega}\tau-\epsilon_{10}\sin\dot{\omega}\tau.
\end{align}

 Indeed, for sufficiently small values of $\dot{\omega}\tau$, the equations (\ref{eq:LL_evolution}) {can be written as}
\begin{align}
\epsilon_{1} (t) & =\epsilon_{10} + \epsilon_{20} \; \dot{\omega}\tau\,,\nonumber \\
\epsilon_{2} (t) & =\epsilon_{20} - \epsilon_{10} \; \dot{\omega}\tau\,.\label{eq:LL_Secular_Variation}
\end{align}
The above equations are identical to the expressions for $\epsilon_{1} (t)$ and $\epsilon_{2}(t)$ present in 
the {\ttfamily ELL1} model. This is because it is straightforward to show that $\dot \omega$ is given either 
by $ \dot \epsilon_1/ \epsilon_{20}$ or $ -\dot \epsilon_2 / \epsilon_{10}$ with the help of 
equations (A14) and (A15) of \cite{Lange2001} while equating $\dot e=0$. 

The above arguments demonstrate that the {\ttfamily ELL1} timing model, detailed in \cite{Lange2001}, accounts 
only for linear-in-time variations of the LL parameters. This approximation may not be appropriate 
while pursuing timing campaigns spanning decades such that $\tau \gtrsim \tau_{\dot\omega} \equiv {2\pi}/{\dot{\omega}}$. 
Additionally, we have included  the $\frac{-3}{2}\, \frac{x}{c}\, \epsilon_1$ term in {our} expression for the \Romer delay.
In contrast, this term was neglected in \cite{Lange2001} as well as  its {\tempo} \citep{Edwards2006_TEMPO2} and \texttt{TEMPO} implementations. 
This term turned out to be crucial while invoking equations (\ref{eq:LL_evolution}) for evolving the LL parameters 
of systems with significant periastron advance.


 However, we would like to emphasize that the {\ttfamily ELL1} model should be quite appropriate to fit the observed 
 TOAs when the span of the timing campaign is much smaller than $\tau_{\dot\omega}$. 
To demonstrate this, we write the barycentric arrival time of a pulse as 
\begin{equation}
t_{\text{obs}}=t_{\text{em}}+\Delta_{R}\,.
\end{equation}
For nature's perfect clocks, we have 
\begin{equation}
t_{\text{em}}=t_{\text{em}}^{(0)}+\frac{N}{f}\,,
\label{eq:tem1}
\end{equation}
where 
$t_{\text{em}}^{(0)}$ is the time of emission of a `reference' pulse,
$f$ is the pulsar rotation frequency, and
$N$ is some integer while ignoring any
variations in pulsar frequency $f$.
The expression for the \Romer delay, namely equation (\ref{eq:Romer_Delay_ELL1}), 
may be written with the help of equations (\ref{eq:LL_Secular_Variation}) as  
\begin{align}
\Delta_{R} & = \frac{x}{c}\left[\sin\Phi+\frac{1}{2}\left(\left(\epsilon_{20}+\dot{\epsilon}_{2}\tau\right)\sin2\Phi-\left(\epsilon_{10}+\dot{\epsilon}_{1}\tau\right)\cos2\Phi\right)\right]  \nonumber\\
  &\quad - \frac{3}{2}\frac{x}{c}\epsilon_{1}  
\nonumber\\
 & = \Delta_{R}^{(\text{\ttfamily ELL1})}-\frac{3}{2}\frac{x}{c}\epsilon_{1}\,,
\label{eq:ELL1K_linear_variations}
\end{align}
where we used the relations 
$\dot{\epsilon}_1=\epsilon_{20}\dot{\omega}$ and 
$\dot{\epsilon}_2=-\epsilon_{10}\dot{\omega}$.  
In the above equation, 
$\Delta_{R}^{(\text{\texttt{ELL1}})}$ stands for the \Romer delay  expressions present in the 
{\ttfamily ELL1} model \citep{Lange2001} that incorporate only linear-in-time  variations  for the Laplace-Lagrange parameters.
The expression for 
the \Romer delay  now becomes 
\begin{align}
\Delta_{R} &= \Delta_{R}^{(\text{\texttt{ELL1}})}-\frac{3}{2}\frac{x}{c}\epsilon_{10}-\frac{3}{2}\frac{x}{c}\epsilon_{20}\dot{\omega}\left(t_{\text{em}}^{(0)}+\frac{N}{f}-T_{\ascnode}\right)\,,
\end{align}
as $\tau=t-T_{\ascnode}$. 
This allows us to write the barycentric arrival times of pulses as
{\small
\begin{align}
t_{\text{obs}} 	&= t_{\text{em}}^{(0)}+\frac{N}{f}+\Delta_{R}^{(\text{\texttt{ELL1}})}-\frac{3}{2}\frac{x}{c}\epsilon_{10}-\frac{3}{2}\frac{x}{c}\epsilon_{20}\dot{\omega}\left(t_{\text{em}}^{(0)}+\frac{N}{f}-T_{\ascnode}\right)\nonumber \\
 				&= t_{\text{em}}^{(0)}+\frac{N}{f}\left(1-\frac{3}{2}\frac{x}{c}\epsilon_{20}\dot{\omega}\right)+\Delta_{R}^{(\text{\texttt{ELL1}})} -\frac{3}{2}\frac{x}{c}\left(\epsilon_{10}+\epsilon_{20}\dot{\omega}\left(t_{\text{em}}^{(0)}-T_{\ascnode}\right)\right) .  
\label{eq:TOA_modfreq}
\end{align}}
A close inspection reveals that the 
last term in equation (\ref{eq:TOA_modfreq}) is a constant over time 
and can be neglected. Additionally, the effect of the `extra term' can be absorbed into the pulsar frequency by redefining it as
\begin{equation}
\label{eq:f_fp}
f'=f\left(1-\frac{3}{2} \frac{x}{c} \epsilon_{20} \dot{\omega}\right)^{-1}.
\end{equation}
The above conclusion holds even if we 
include $\dot{f}$ in equation (\ref{eq:tem1}).
A straightforward calculation allows us to 
re-write  equation (\ref{eq:TOA_modfreq})  as
\begin{equation}
t_{\text{obs}}=t_{\text{em}}^{(0)}+\frac{N}{f'}-\frac{\dot{f}'N^{2}}{2f'^{3}}+\Delta_{R}^{(\texttt{ELL1})}\,,
\end{equation}
where we neglected certain constant terms and 
\begin{equation}
\dot{f}'=\dot{f}\left(1-\frac{3}{2}\frac{x}{c}\epsilon_{20}\dot{\omega}\right)^{-2}.
\end{equation}

The above arguments demonstrate that the {\ttfamily ELL1} timing model should be sufficient to analyze timing data, provided 
$\tau \ll \tau_{\dot\omega}$ for ensuring the validity of equations (\ref{eq:LL_Secular_Variation}).
However, the model may not be appropriate  while pursuing long-term timing campaigns of short orbital
period pulsar binaries that can accommodate, in principle, non-negligible periastron advance.
It should be noted that the 
higher order corrections to equation (\ref{eq:LL_Secular_Variation}) 
may be partially absorbed into higher derivatives of $f$, 
namely $\dot{f}$, $\ddot{f}$ etc., provided 
$\tau \ll \tau_{\dot\omega}$. This may be relevant during the coherent timing of pulsar binaries like 
SAX J1808.4$-$3658 \citep{Patruno2012}.
However,  it may not be desirable to treat temporal variations in the LL parameters  to be linear-in-time 
while pursuing long-term timing campaigns of short-period binary pulsars.
This is what we pursue in the next section.

\section{\texttt{ELL1\lowercase{k}} model and its observational implications }
\label{sec:measurability}
 To probe the observational consequences of {our} equations 
 (\ref{eq:Romer_Delay_ELL1}) and
  (\ref{eq:LL_evolution}), we implemented a modified 
 version of the  {\ttfamily ELL1} timing model in \texttt{TEMPO2}.
 This model, available in the latest version of \tempo, is referred to as 
  {\ttfamily ELL1k} 
 due to the use 
 of $k$ to represent the dimensionless fractional periastron advance per orbit \citep{DS1988}.
 The relevant expressions, required to implement the {\ttfamily ELL1k} timing model, 
 are listed in  Appendix \ref{sec:t2_inputs} and they do contain contributions from the above mentioned 
 additional term.
 To probe the relevance of {our improvements}, 
 we simulate binary pulsar TOAs using {our}  {\ttfamily ELL1k} model  
 for different observation spans and white timing noise amplitudes using the {\ttfamily fake} plug-in of {\ttfamily TEMPO2}. 
 The resulting TOAs are fitted using 
 the 
 {\ttfamily ELL1} model.
 We explore the goodness of the fit by varying $\dot\omega$ values,
  observation spans and the amplitude of the white timing noise.
If the fit exhibits systematic variations in the residuals, we conclude that 
our {\ttfamily ELL1k} model  
should be relevant for analyzing TOAs from such binary pulsar configurations.
Additionally, we explore the implications of red (i.e., non-Gaussian) timing noise in these simulations 
and our observations are summarized in Section \ref{sec:RedNoise}.
We begin  by describing why the long-term timing of a recently discovered system should be interesting 
 from our point of view. 

\subsection{Diamond planet-pulsar binary as a possible test bed for {our}  {\ttfamily ELL1k} timing model}
\label{J1719-1438}
The recently discovered 
PSR J1719$-$1438  is a unique binary millisecond pulsar with an ultra-low mass/planetary companion having an 
orbital period of about 2.18 hours and eccentricity of $\sim 8\times 10^{-4}$ \citep{Bailes2011}.
We list in  Table \ref{tab:J1719-1438_ephem} a few relevant 
orbital parameters, extracted from  \cite{Ng2014_Timing}. 
Accurate timing enabled 
\cite{Bailes2011}  to demonstrate that  
the mass  of the  companion should be 
 greater than $0.0011292M_{\astrosun}$ (about $1$ Jupiter mass) while 
 its radius is constrained to be less than $28432.9\text{ km}$ (about 40\% of Jupiter radius). 
\begin{table}
\begin{tabular}{|c|c|c|}
\hline 
Parameter & Unit & Value\tabularnewline
\hline 
\hline 
$P_{b}$ & days & 0.09070629\tabularnewline
\hline 
$x$     & lt-s & 0.0018212\tabularnewline
\hline 
$T_{\ascnode}$ & MJD &  55235.516505 \tabularnewline
\hline 
$\epsilon_1$ &  &  $-7\times 10^{-4}$\tabularnewline
\hline 
$\epsilon_2$ &  &  $4\times 10^{-4}$\tabularnewline
\hline 
\end{tabular}
\caption{Orbital parameters of PSR J1719$-$1438, obtained using the {\ttfamily ELL1} timing model, 
listed in \citet{Ng2014_Timing}.
}
\label{tab:J1719-1438_ephem}
\end{table}

 It turns out that this binary system should exhibit, in principle, the advance of periastron.
 For PSR J1719$-$1438, 
the advance of periastron should arise due to the dominant order general relativistic, static tidal and classical spin-orbit (SO) interactions.
The relevant expressions for $\dot{\omega}_{\text{GR}}$, $\dot{\omega}_{\text{tidal}} $ and $\dot{\omega}_{\text{SO}} $ are given by 
(see, e.g., \citet{Willems2008})
\begin{subequations}
\label{eq:omdot}
\begin{align}
\dot{\omega}_{\text{GR}} &= \frac{3}{1-e^{2}}\left(\frac{GM}{c^{3}}\right)^{2/3}\left(\frac{2\pi}{P_{b}}\right)^{5/3}\,,
\label{eq:omdot_GR}\\
\dot{\omega}_{\text{tidal}} & =15\left(\frac{R_{c}}{a}\right)^{5}\left(\frac{m_{p}}{m_{c}}\right)\frac{2\pi}{P_{b}}k_{2}\left(\frac{1+\frac{3}{2}e^{2}+\frac{1}{8}e^{4}}{\left(1-e^{2}\right)^{5}}\right) \,,
\label{eq:omdot_tidal}\\
\dot{\omega}_{\text{SO}} & =\frac{2\pi}{P_{b}}\left(\frac{R_{c}}{a}\right)^{5}\left(\frac{M}{m_{c}}\right)\left(\frac{P_{b}}{P_{c}}\right)^{2}\frac{k_{2}}{\left(1-e^{2}\right)^{2}}\,, 
\label{eq:omdot_LS}
\end{align}
\end{subequations}
where 
$R_c$, $P_c$ and $k_2$ are respectively the radius, rotational period and quadrupolar tidal Love number 
(also known as quadrupolar apsidal motion constant) of the companion.

  It should be obvious that general relativistic contributions are {well-constrained} by the 
  existing observations \citep{Ng2014_Timing}. 
  However, additional assumptions will be required to estimate classical contributions to $\dot \omega$.
We may estimate a value for the semi-major axis \good{of the relative orbit} $a$ to be $2.1789239$ lt-s by   
assuming an edge-on orbit and a neutron star mass $m_p=1.35 M_{\astrosun}$ \citep{Manchester1993_ATNF_Catalog}.
Unfortunately, it is rather difficult to provide a firm estimate for $k_2$ as we do not know the exact nature of the companion.
It is reasonable to expect that the companion's $k_2$ may lie between the following two possible ranges.
In case the companion is a low-mass white dwarf, $k_2$ can be between 
 $\sim 0.01$ to $\sim 0.1$ \citep{Valsecchi+2012}.
 Possible $k_2$ values can be 
from $\sim 0.001$ to $\sim 0.01$ for main sequence stars \citep{Stothers1974} while 
 $k_2$ value can be as high as $\sim 0.5$ for the Jovian planets \citep{Wahl2016}.
We choose a rather conservative estimate for the companion's apsidal motion constant and let 
$k_2=0.01$,
and impose co-rotation of the companion with its orbital motion. This leads to 
\begin{align}
\dot{\omega}_{\text{GR}} 	& \simeq 13.3\degree\text{/yr}\,,\nonumber\\
\dot{\omega}_{\text{tidal}}	& \simeq 40.5\degree\text{/yr}\,,\nonumber\\
\dot{\omega}_{\text{SO}} 	& \simeq 2.7\degree\text{/yr}\,,
\label{eqn:omega}
\end{align}
and the total $\dot{\omega} \simeq 56.5\degree\text{/yr}$ for PSR J1719$-$1438.

  It is important to note that our $\dot{\omega}_{\text{tidal}}$ and $\dot{\omega}_{\text{SO}}$ estimates are extremely 
sensitive to the ratio $R_{c}/{a}$, as evident from 
equations (\ref{eq:omdot_tidal}) and (\ref{eq:omdot_LS}).
For example, if the actual $R_{c}$ value 
is 50\% of the current upper bound, $\dot{\omega}_{\text{tidal}}$ and $\dot{\omega}_{\text{SO}}$ will reduce by a factor of $2^{5}=32$ from the above listed values. 
This implies that our possible estimates for  $\dot{\omega}_{\text{tidal}}$ and $\dot{\omega}_{\text{SO}}$  should be 
treated with a high degree of caution.
However, our $\dot{\omega}_{\text{GR}}$ estimate should be fairly accurate due to
its crucial dependence on the accurately measured $P_{b}$ value.
Additionally, the inferred total mass of the binary system 
is expected to be close to the typical pulsar mass, namely $1.35M_{\astrosun}$ as
observations strongly suggest the presence of an ultra-low mass companion in PSR J1719$-$1438 \citep{Bailes2011}.
Therefore, we employ our $\dot{\omega}_{\text{GR}}$ estimate while exploring the 
observational implications of {the} {\ttfamily ELL1k} model for the PSR J1719$-$1438 binary system.

\begin{figure}
\label{fig:J1919-1438_2yr}
\begin{subfigure}{.5\textwidth}
	\centering	
	\includegraphics[scale=0.7]{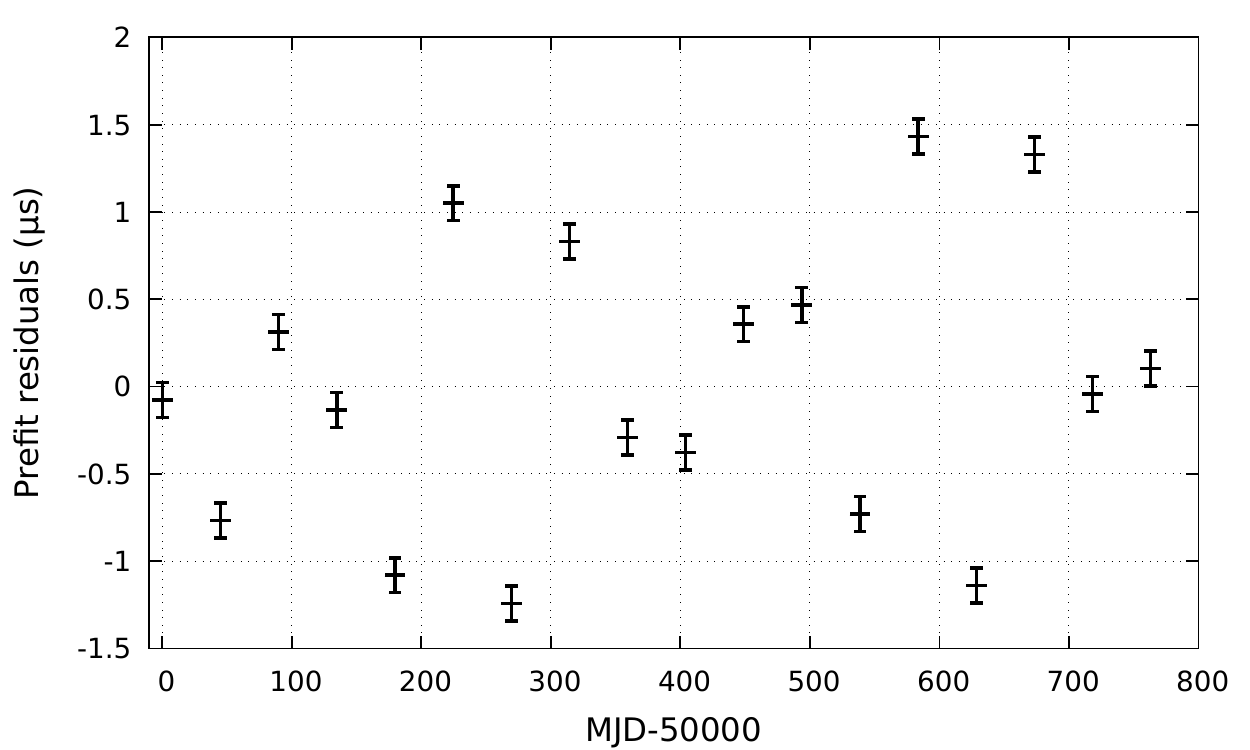}
	\caption{Pre-fit}
	\label{fig:J1919-1438_2yr-prefit}
\end{subfigure}
\begin{subfigure}{.5\textwidth}
	\centering	
	\includegraphics[scale=0.7]{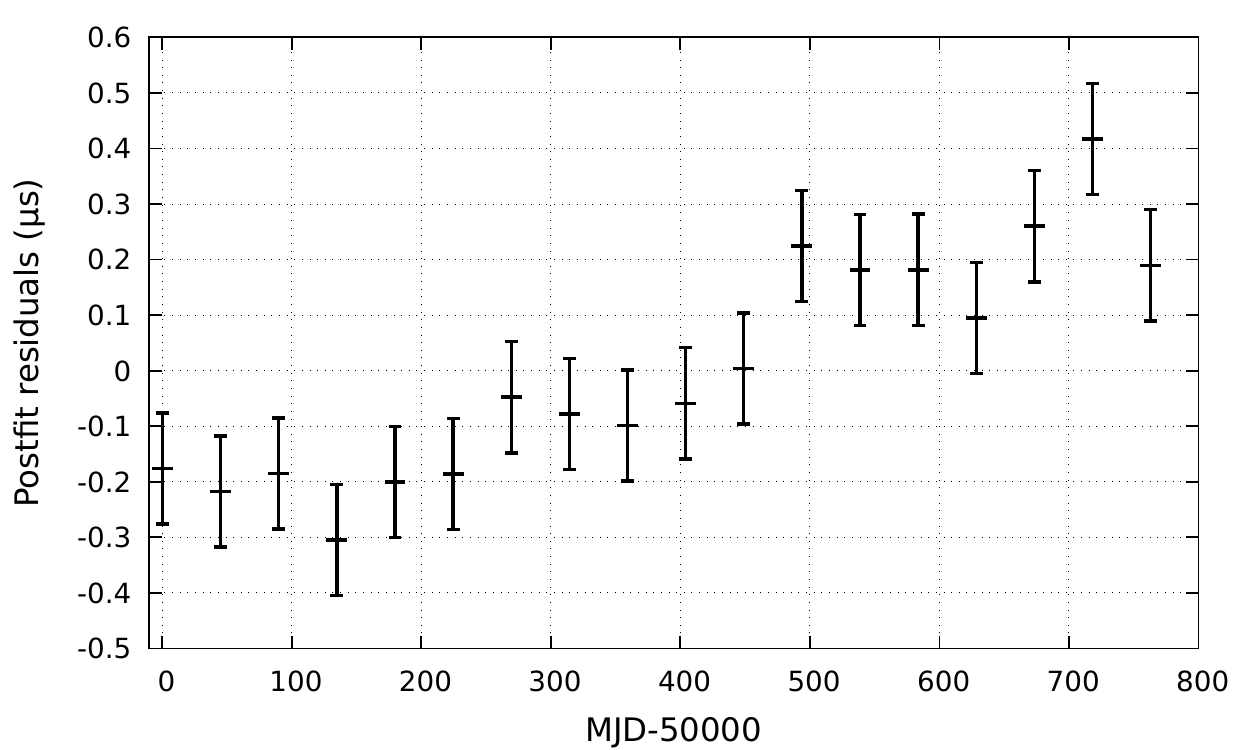}
	\caption{After fitting for $\epsilon_{1}$, $\epsilon_{2}$, ${\dot\epsilon}_{1}$, ${\dot\epsilon}_{2}$}
	\label{fig:J1919-1438_2yr-after_epsdot}
\end{subfigure}
\begin{subfigure}{.5\textwidth}
	\centering	
	\includegraphics[scale=0.7]{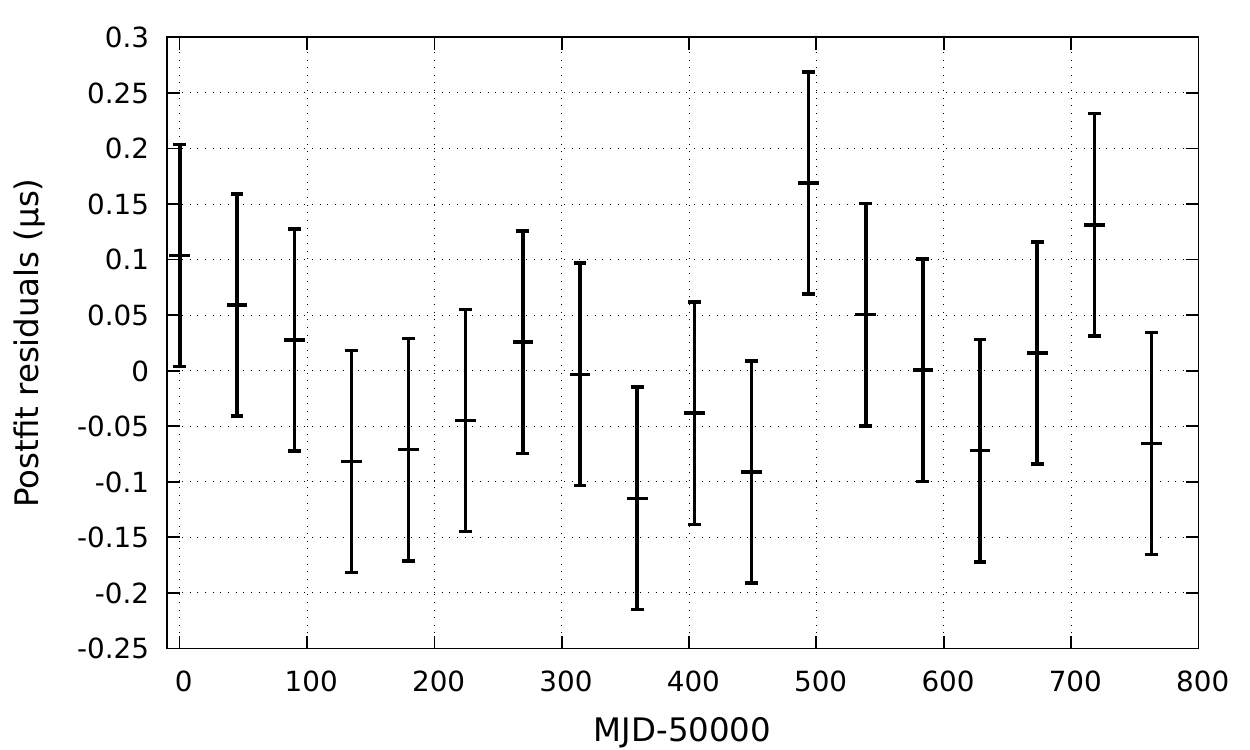}
	\caption{After additionally fitting for $f$ (Reduced $\chi^2=0.92$)}
	\label{fig:J1919-1438_2yr-after_f0}
\end{subfigure}
\caption{
{\tempo} simulations of PSR J1719$-$1438 spanning two years duration. For these simulations, TOAs were generated using the \texttt{fake} plug-in 
of {\tempo} with the {\ttfamily ELL1k} timing model for parameters listed in Table \ref{tab:J1719-1438_ephem} and we let 
$\dot \omega$ 
 take its GR value, namely $13.3^{\circ }/\text{yr}$. 
The top panel (a) shows the pre-fit timing residuals to our data where the fit was done using 
the {\ttfamily ELL1} model for Table \ref{tab:J1719-1438_ephem} parameters.
The middle panel (b) shows the post-fit residuals after fitting $\epsilon_{1}$, $\epsilon_{2}$, ${\dot \epsilon}_{1}$ and 
${\dot \epsilon}_{2}$ of the  {\ttfamily ELL1} model. The bottom panel (c) shows the post-fit residuals after additionally fitting for the pulsar spin frequency $f$.
The bottom panel shows that $\dot \omega$ induced variations can be absorbed into the \emph{unknown} pulsar frequency if 
$\tau \ll \tau_{\dot \omega}$.
}
\end{figure}

\begin{figure}
\begin{subfigure}{.5\textwidth}
	\centering	
	\includegraphics[scale=0.7]{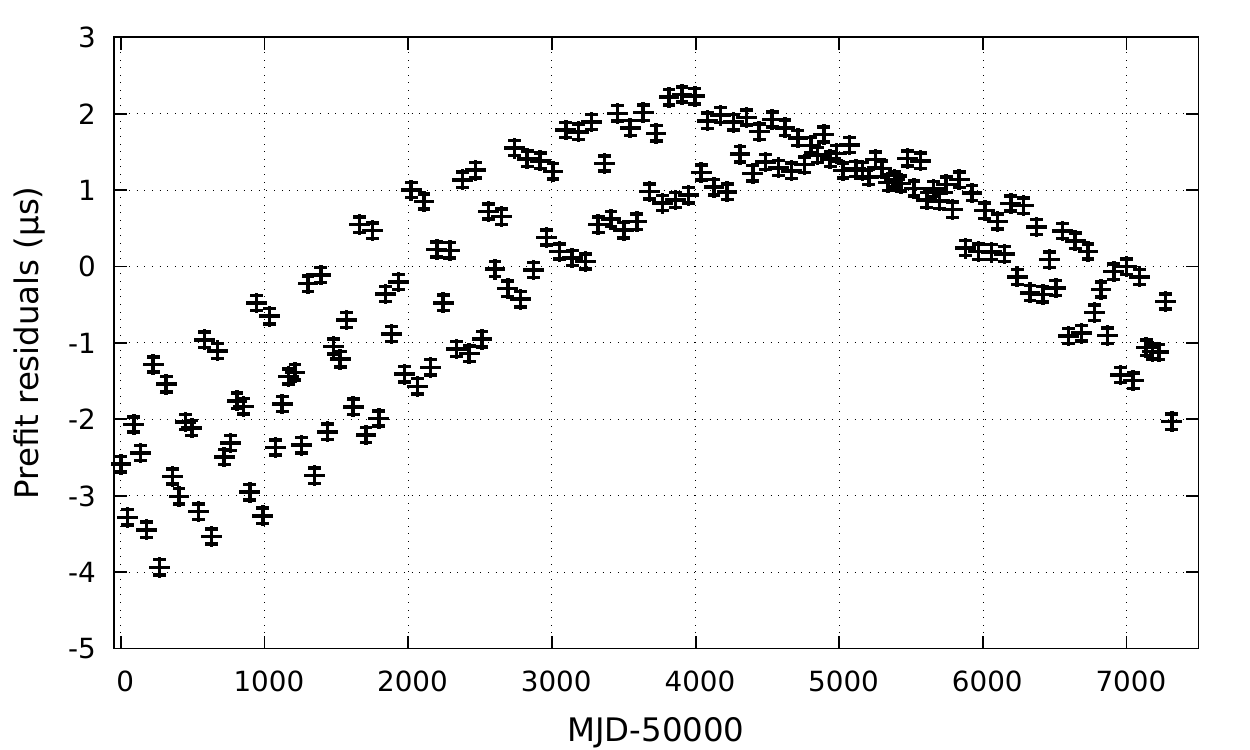}
	\caption{Pre-fit}
	\label{fig:J1919-1438_20yr-prefit}
\end{subfigure}
\begin{subfigure}{.5\textwidth}
	\centering	
	\includegraphics[scale=0.7]{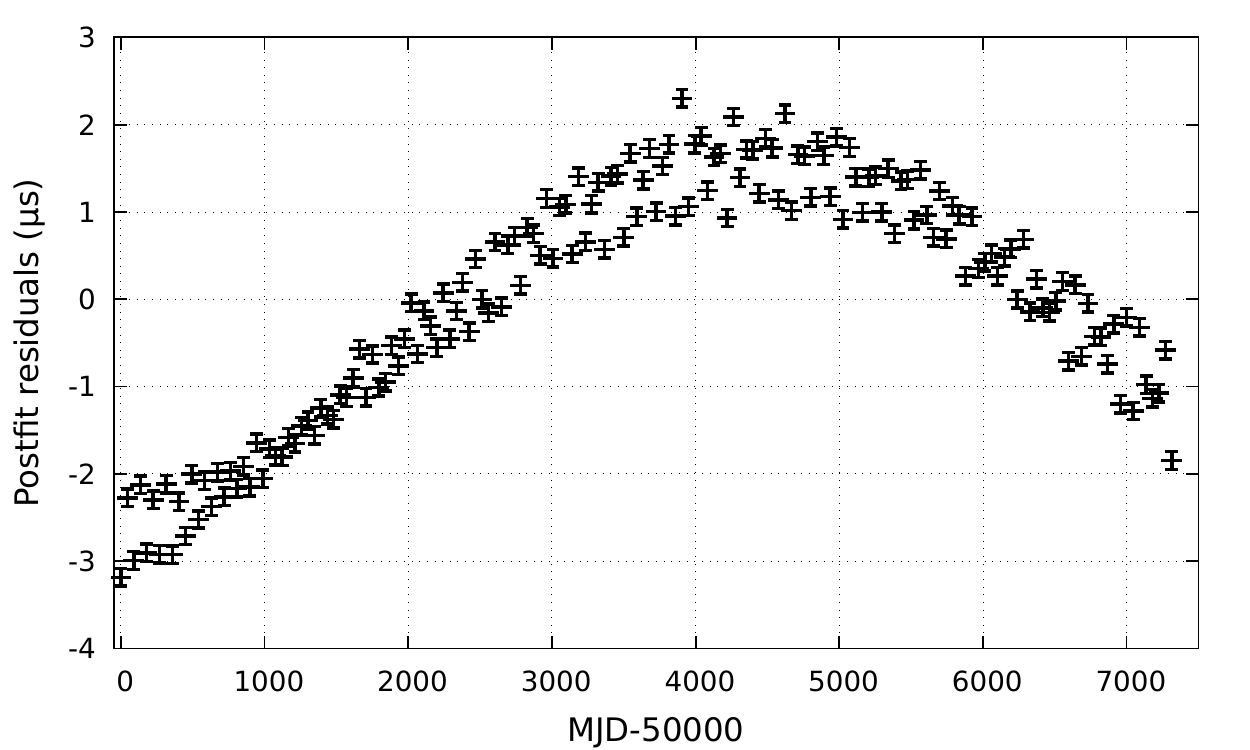}
	\caption{After fitting for $\epsilon_{1}$, $\epsilon_{2}$, ${\dot\epsilon}_{1}$, ${\dot\epsilon}_{2}$}
	\label{fig:J1919-1438_20yr-after_epsdot}
\end{subfigure}
\begin{subfigure}{.5\textwidth}
	\centering	
	\includegraphics[scale=0.7]{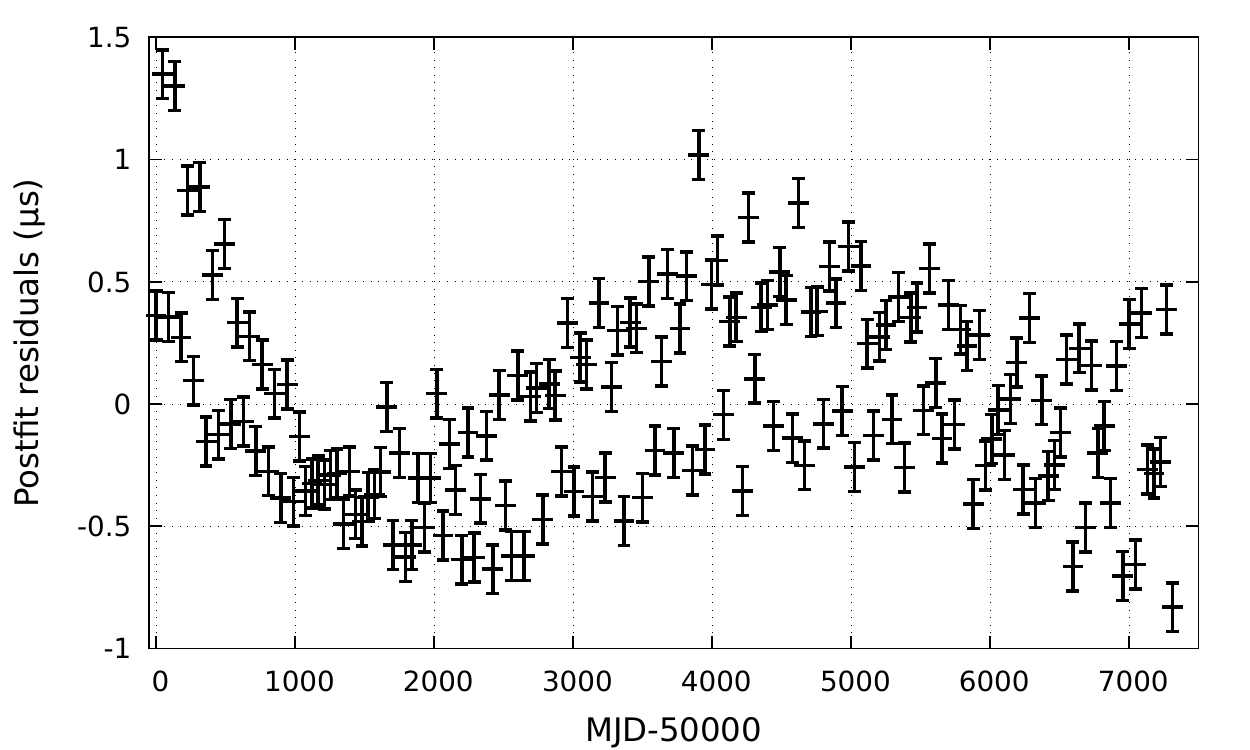}
	\caption{After additionally fitting for $f$ and $\dot{f}$ (Reduced $\chi^2=16.57$)}
	\label{fig:J1919-1438_20yr-after_fdot}
\end{subfigure}
\caption{We repeat the {\tempo} analysis of Figures \ref{fig:J1919-1438_2yr-prefit}-\ref{fig:J1919-1438_2yr-after_f0} for a time span of 20 years.
Figure \ref{fig:J1919-1438_20yr-after_fdot} shows that the general evolution of the LL parameters can not be 
\good{accounted} for by the 
{\ttfamily ELL1} timing model even after fitting for $f$ and its time derivative $\dot f$. 
}
\end{figure}

\begin{figure}
\begin{subfigure}{.5\textwidth}
	\centering	
	\includegraphics[scale=0.7]{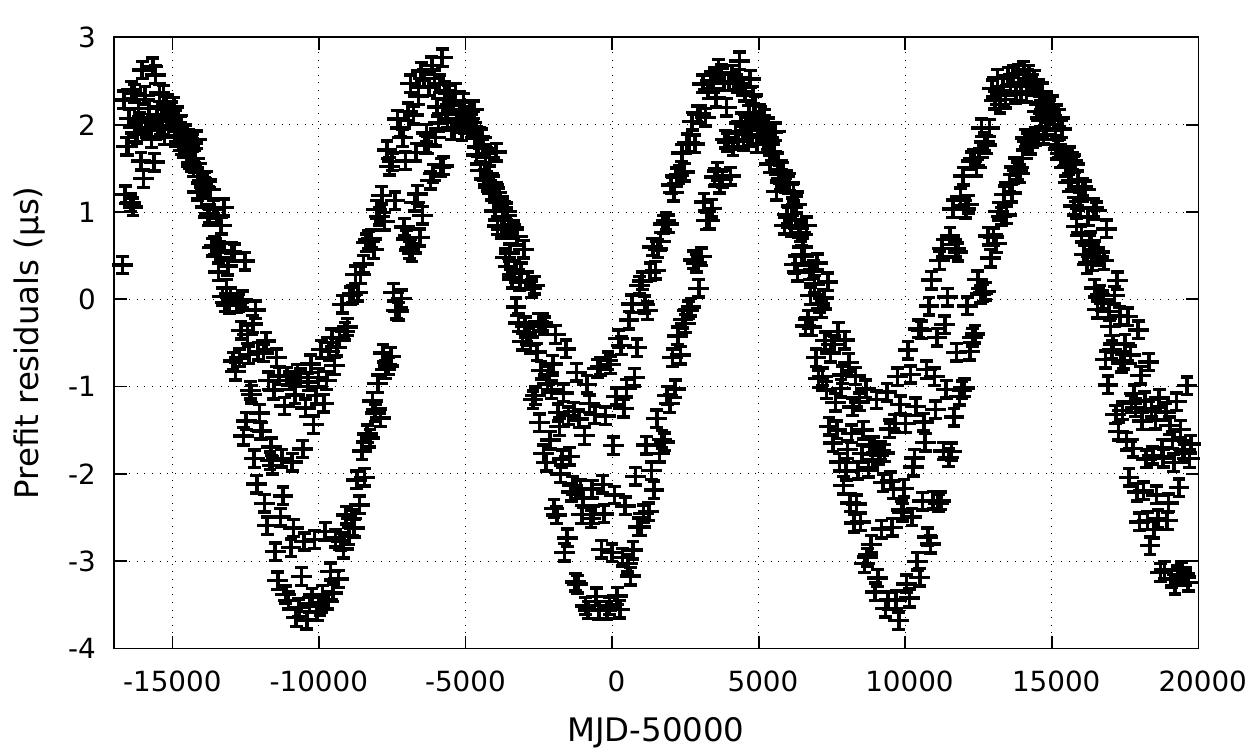}
	\caption{Pre-fit}
	\label{fig:J1919-1438_100yr-prefit}
\end{subfigure}
\begin{subfigure}{.5\textwidth}
	\centering	
	\includegraphics[scale=0.7]{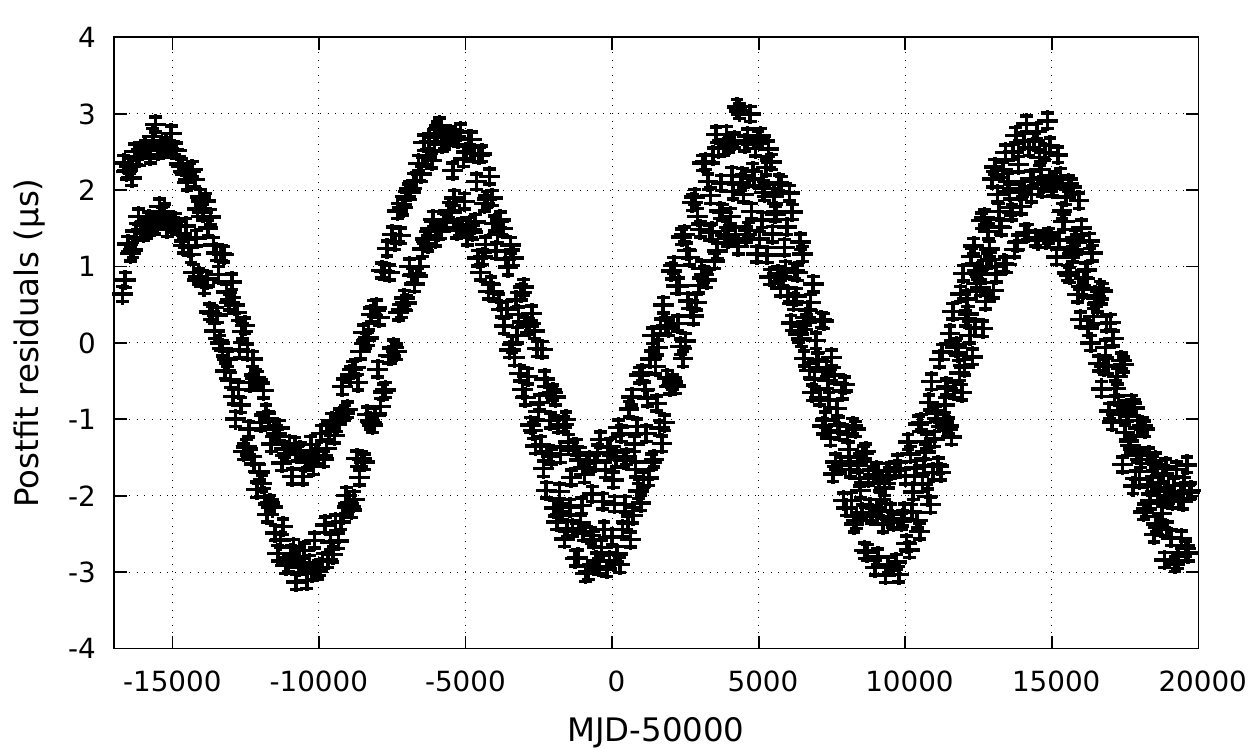}
	\caption{After fitting for $\epsilon_{1}$, $\epsilon_{2}$, ${\dot\epsilon}_{1}$, ${\dot\epsilon}_{2}$}
	\label{fig:J1919-1438_100yr-after_epsdot}
\end{subfigure}
\begin{subfigure}{.5\textwidth}
	\centering	
	\includegraphics[scale=0.7]{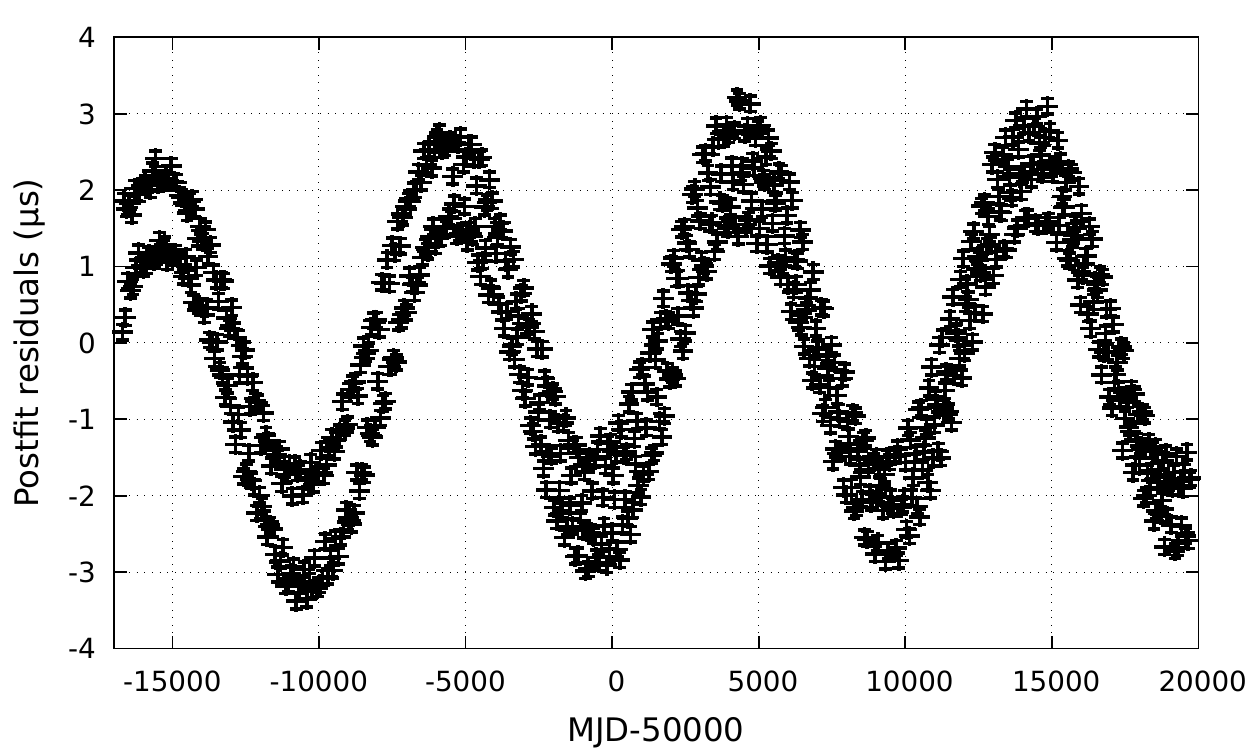}
	\caption{After fitting additionally for $f$ and $\dot{f}$ (Reduced $\chi^2=277.55$)}
	\label{fig:J1919-1438_100yr-after_fdot}
\end{subfigure}
\caption{We repeat the {\tempo} simulations for PSR J1719$-$1438 for a time window of 100 years. 
The three plots were obtained similarly to the Figures \ref{fig:J1919-1438_2yr-prefit}-\ref{fig:J1919-1438_2yr-after_f0}.
The effect of periastron advance is impossible to miss in the residuals  while timing PSR J1719$-$1438  using the existing
{\ttfamily ELL1} timing model.
} 
\end{figure}

 We display in three figures our 
 simulations for the above binary that spans 
$2$ years (Figures \ref{fig:J1919-1438_2yr-prefit}-\ref{fig:J1919-1438_2yr-after_f0}), 
$20$ years (Figures \ref{fig:J1919-1438_20yr-prefit}-\ref{fig:J1919-1438_20yr-after_fdot}) and 
$100$ years (Figures \ref{fig:J1919-1438_100yr-prefit}-\ref{fig:J1919-1438_100yr-after_fdot}).
For these plots, we simulated 
\good{one} observation \good{each} in every $45$ days
and chose a white timing noise amplitude of $100$ ns while assuming  
the absence of any red timing noise.
Timing residuals arise as we employ the standard {{\ttfamily ELL1}} timing model to analyze the TOAs generated 
with {the} {{\ttfamily ELL1k}} model.
Plots in Figures \ref{fig:J1919-1438_2yr-prefit}-\ref{fig:J1919-1438_2yr-after_f0} demonstrate that
it will not be possible to identify the effects of $\dot \omega$ 
if the total duration of the observation is too short compared to the periastron advance timescale, namely $\tau_{\dot\omega} = 2\, \pi/ \dot \omega$, 
which in this case is $\sim 27$ years while assuming only the GR contribution.
Clearly, any systematics arising from our proposed modifications
 are absorbed in $f$ as was shown in Section \ref{sec:Advance-of-periapsis_ELL1}. 
However, 
the effect of $\dot{\omega}$ manifests as systematic variations in the post-fit residuals
when span of the observations is comparable to  $\tau_{\dot\omega}$ as evident from plots in 
Figures \ref{fig:J1919-1438_20yr-prefit}-\ref{fig:J1919-1438_20yr-after_fdot}.
This point is further emphasized in Figures \ref{fig:J1919-1438_100yr-prefit}-\ref{fig:J1919-1438_100yr-after_fdot}  
which show the case where the total observation duration is much greater than the periastron advance timescale for 
the PSR J1719$-$1438 binary system.

\good{}

The inferences, gleaned from these figures, are  
clearly consistent with Section \ref{sec:Advance-of-periapsis_ELL1} conclusions. 
The linear variation in the timing residuals, present in 
Figure \ref{fig:J1919-1438_2yr-after_epsdot}  is fully expected due to 
the additional term in 
our equation (\ref{eq:ELL1K_linear_variations}).
In comparison, the periodic nature of the timing residuals present in 
 Figures \ref{fig:J1919-1438_100yr-prefit}-\ref{fig:J1919-1438_100yr-after_fdot} 
 naturally arise from the `exact' temporal evolutions of 
  the Laplace-Lagrange parameters, namely equation (\ref{eq:LL_evolution}).
These 
simulations open up the tantalizing possibility of measuring  the effects of  
$\dot\omega$ in the PSR J1719$-$1438 binary system, 
provided high-cadence 10 year timing data is available.
It is possible that the effects of $\dot\omega$ may barely be detectable now 
as the system was discovered in 2011 \citep{Bailes2011}, especially if there exists significant contributions 
from the tidal and classical spin-orbit interactions.
Obviously, this requires the availability of high-precision TOAs and 
the absence of significant red timing noise.
The present 
simulations should provide sufficient motivation 
to pursue  long-term, high-cadence timing of  PSR J1719$-$1438.
\good{The cadence of an observation campaign for measuring $\dot\omega$ should be such that it samples a cycle of the periastron (with period $\tau_{\dot\omega}$) at a sufficient rate. We have checked that the systematic timing residuals, visible in Figures \ref{fig:J1919-1438_20yr-prefit}-\ref{fig:J1919-1438_20yr-after_fdot} and \ref{fig:J1919-1438_100yr-prefit}-\ref{fig:J1919-1438_100yr-after_fdot}, persist even when the cadence is as sparse as one observation per year.}
Finally, we would like to note that the 100 ns TOA uncertainties,
invoked in our simulations, are usually achievable by the existing PTA telescopes.

 We observe that the extra term, namely  $-3x\epsilon_1/2c$ in equation (\ref{eq:Romer_Delay_ELL1}),  contributed significantly to the 
 long-term oscillatory behavior of the timing residuals as visible in Figures \ref{fig:J1919-1438_20yr-prefit} and \ref{fig:J1919-1438_100yr-prefit}.
In the absence of this extra term, the resulting timing residuals turned out to be smaller in magnitudes and their
long timescale variations were similar to what we display in Figure \ref{fig:J1919-1438_2yr-HW}.
This suggests that the additional term should be relevant while extracting the effect of periastron advance 
from the timing of binary pulsars like PSR J1719$-$1438.
Additionally, the extra term forces the observed spin frequency to be different from its intrinsic one 
as evident from equation (\ref{eq:f_fp}). This may have implications while searching for
continuous gravitational waves from pulsars in binaries \citep{Watts2008}.
This is because continuous gravitational waves from such systems are expected at $2 f$ rather 
than at $2f'$ and further investigations will be required to quantify its relevance.

 An additional point that requires further investigation is 
 related to the inclusion of Shapiro delay.
 In general, the Shapiro delay is degenerate with the 
 \Romer delay \citep{Lange2001,Freire2010}.
This ensures that   
the measured $x$ and $\epsilon_{10}$ values are  different from their intrinsic values. However, our
 extra term in the \Romer delay can, in principle, break this degeneracy, provided the observation span is sufficiently large
 and the  system exhibits significant periastron advance.
This interesting possibility may be relevant for systems
 like PSR J0348+0432 \citep{afw+13} and demands further studies.
 \good{This may require incorporating $\dot\omega$ in an exact manner in the \texttt{ELL1H} model of \citet{Freire2010}, which expresses Shapiro delay as a Fourier series in the orbital phase $\Phi$ for low-eccentricity, low to moderate-inclination binaries.}
In the next subsection, we explore possible binary pulsar systems where the $\dot\omega$ effect may be 
\good{measurable} in the near future.


\subsection{Are there additional {\ttfamily ELL1k}-relevant pulsar binaries?}

 A close inspection of  equation (\ref{eq:omdot}) reveals
that pulsar binaries with short orbital periods and non-compact companions should be interesting 
candidates for the  {\ttfamily ELL1k} timing model. 
It turns out that there exist 
a number of  tiny-$e$ binaries 
which can, in principle, exhibit non-negligible $\dot \omega$ as evident  from 
the current ATNF pulsar catalog\good{\footnote{Available at \url{http://www.atnf.csiro.au/research/pulsar/psrcat/}}} \citep{Manchester1993_ATNF_Catalog}.
In Table \ref{tab:omdot_pulsars}, we display the most promising  {\ttfamily ELL1k}-relevant  pulsar binaries present in the catalog for which the
GR-induced periastron advance timescale is less than 500 years.
Unfortunately, it is rather difficult to provide plausible estimates for the 
$\dot\omega_{\text{tidal}}$ and $\dot\omega_{\text{SO}}$ contributions in many of these interesting 
candidates.
This is expected due to our ignorance about the likely values for the companion mass $m_c$, its radius $R_c$ and the 
associated Love number $k_2$.

 Timing of tiny eccentric binary pulsars usually yields an estimate for the binary mass function in the absence of 
 any measurable Shapiro delay. Therefore, an estimate for the companion mass requires us to impose edge-on orbital 
 orientation and let the pulsar mass  be $m_p=1.35\,M_{\astrosun}$. 
 This arises from the definition of the mass function 
\begin{equation}
f(m_{c})=\frac{4\pi^{2}}{G}\frac{\left(a_{p}\sin\iota\right)^{3}}{P_{b}^{2}}=\frac{\left(m_{c}\sin\iota\right)^{3}}{M^{2}}\,.
\end{equation}
Measurement of the \Romer delay provides an estimate for $a_p\, \sin \iota $ and $P_b$ which allows us to 
estimate 
$m_c$ with the above assumptions. This is how we obtain the $m_c$ values listed in Table \ref{tab:omdot_pulsars}.

To obtain an estimate for the companion's radius, we need to specify its nature.
If the companion is believed to be a white dwarf, we invoke the  well-known white dwarf mass-radius relation to compute $R_c$, given $m_c$. 
Specifically, we employ the following  numerical approximation to the  white dwarf mass-radius relation \citep{Carvalho2015}
\begin{equation}
\frac{m}{M_{\astrosun}} = \frac{\alpha R + \beta}{e^{\gamma R^2}+\delta}\,,
\label{eq:WD_MR}
\end{equation} 
where 
$\alpha = 2.325\times 10^{-5} \text{ km}^{-1}$, 
$\beta = 0.4617$, 
$\gamma = 7.277 \times 10^{-9} \text{ km}^{-2}$ and 
$\delta = -0.644$.
A caveat is that the white dwarf mass-radius relation may not be appropriate for ultra-low mass companions, and may give unrealistically large radii.
Influenced by \citet{Bailes2011},  we place upper bounds on companion radii by arguing that the companion must be smaller than its Roche lobe.
In such cases,  we invoke the standard  radius of the Roche lobe for ultra-low mass companion \citep{Paczynski1971}
\begin{equation}
R_{L}=0.462 \, a\left(\frac{m_{2}}{M}\right)^{\frac{1}{3}}\,.
\label{eq:Roche_lobe_radius}
\end{equation}
If the companion radius given by equation (\ref{eq:WD_MR}) is larger than the above upper bound, given by equation (\ref{eq:Roche_lobe_radius}), 
we employ  the lower $R_c$ estimate  in our calculations. We highlight such cases by using {\bf boldface} font for $R_c$ in Table \ref{tab:omdot_pulsars}.
The third category involves the possibility that pulsar companions can be 
main sequence stars. In this case,  we employ the following  empirical 
 main sequence mass-radius relation, relevant for low mass stars, to estimate the companion radius \citep{Demircan1991}.
\begin{equation}
\frac{R}{R_{\astrosun}} = 1.06 \left(\frac{m}{M_{\astrosun}}\right)^{0.945} 
\,.
\end{equation}
These cases are marked by the star ($\star$) symbol in the $R_c$ column of Table \ref{tab:omdot_pulsars}.

 The third parameter, required to obtain $\dot{\omega}_{\text{tidal}}$ and $\dot{\omega}_{\text{SO}}$ estimates, 
 is clearly the apsidal motion constant $k_2$. We let  
  $k_2=0.01$ for compact degenerate companions and $k_2=0.001$ for main sequence companions. 
 Unfortunately, it is rather impossible to obtain an estimate for the rotational period of pulsar companions.
 Therefore, we choose $P_c$ to be identical to  the binary  orbital period which requires tidal locking.
Clearly, these assumptions ensure that the listed $\dot{\omega}_{\text{tidal}}$ and $\dot{\omega}_{\text{SO}}$  
values in Table \ref{tab:omdot_pulsars} should be taken as very rough estimates.
 
We gather from  Table \ref{tab:omdot_pulsars} that  there are 
\good{six} pulsars for which $\dot{\omega}$ may be observable within $\sim$50 years
of their discovery. The effect of periastron advance in these systems, 
namely PSRs
J1719$-$1438 \citep{Bailes2011}, 
J0636+5129 \citep{slr+14}, 
J2339$-$0533 \citep{rbs+14,rs11}, 
J2215+5135 \citep{hrm+11}, 
J0348+0432 \citep{lbr+13,afw+13} and 
J0023+0923 \citep{hrm+11},
may be more predominant if they can have non-negligible  $\dot{\omega}_{\text{tidal}}$ and $\dot{\omega}_{\text{SO}}$ contributions.
In particular, J1719$-$1438 and J0636+5129  may well 
have shorter periastron advance timescales (say $< 5$ years), provided they have significant tidal and spin-orbit contributions to $\dot \omega$. 
\good{Unfortunately, J2215+5135 is a redback pulsar for which it is difficult to obtain a coherent timing solution over long time spans \citep{aaa+13} and therefore the \texttt{ELL1k} model may not be relevant for this binary.}
High-cadence timing of these pulsars \good{(except J2215+5135)} should allow us to obtain observationally relevant bounds on $\dot\omega$.
For these pulsars, measurement of $\dot\omega$ should lead to constraints on the companion's radius and/or Love number.


 A number of pulsars listed in Table \ref{tab:omdot_pulsars} are employed in the present PTA 
experiments for detecting nanohertz gravitational waves \citep{vlh+16}.
Fortunately, none of these pulsars have periastron advance timescale less than 100 years, and therefore we do not expect advance of periastron to be relevant for Pulsar Timing Arrays. In Table \ref{tab:omdot_pulsars}, 
PTA pulsars are marked with {\colorbox[HTML]{C0C0C0}{gray background}}. 
In what follows, we invoke a more realistic scenario where \emph{red noise} is present in our simulated TOA measurements to probe the observational feasibility of our above conclusions.

\begin{table*}
\begin{tabular}{|l|l|l|l|r|l|r|r|r|r|r|r|r|r|}
\hline
PSR J        & $P_b$     & $x$        & $e$       & $m_c$               & {\scriptsize Comp.}  & $a$    & $R_c$   & $\dot{\omega}_\text{GR}$ & $\dot{\omega}_\text{tidal}$  & $\dot{\omega}_\text{SO}$  & $\dot{\omega}$    & $\tau_{\dot{\omega}}$ & $\tau_{\dot{\omega}_\text{GR}}$ \\ 
            & {\scriptsize(days)}    & {\scriptsize(lt-s)}   &         & {\scriptsize($10^{-3}$ $M_{\astrosun}$)} &  {\scriptsize Type}       & {\scriptsize($10^3$ km)} & {\scriptsize($10^3$ km)} & {\scriptsize($\degree$/yr)} & {\scriptsize($\degree$/yr)}    & {\scriptsize($\degree$/yr)} & {\scriptsize($\degree$/yr)} & {\scriptsize(yr)}                        & {\scriptsize(yr)}                                    \\ \hline
1719$-$1438                         & 0.090706 & 0.00182 & 8.0E-4 & 1.1                   & UL & 653                   & \textbf{28}                         & 13.3                  & 40.5                  & 2.7                   & 56.5                  & 6.4                   & 27.1                  \\ \hline
0636+5129                         & 0.066551 & 0.00899 & 2.2E-5 & 6.9                   & UL & 532                   & 26                                  & 22.3                  & 17.3                  & 1.2                   & 40.8                  & 8.8                   & 16.1                  \\ \hline
2339$-$0533                         & 0.193098 & 0.61166 & 2.1E-4 & 257.2                 & MS & 1146                  & \cellcolor[HTML]{FFFFFF}204$^\star$ & 4.2                   & 9.7                   & 0.8                   & 14.7                  & 24.5                  & 85.1                  \\ \hline
2215+5135                         & 0.172502 & 0.46814 & 1.1E-5 & 207.9                 & MS & 1052                  & \cellcolor[HTML]{FFFFFF}167$^\star$ & 5.0                   & 7.5                   & 0.6                   & 13.1                  & 27.5                  & 72.0                  \\ \hline
0348+0432                         & 0.102424 & 0.14098 & 2.4E-6 & 83.9                  & He & 723                   & 18                                  & 11.3                  & 0.0                   & 0.0                   & 11.3                  & 31.8                  & 31.9                  \\ \hline
0023+0923                         & 0.138799 & 0.03484 & 2.4E-5 & 16.4                  & UL & 871                   & 24                                  & 6.6                   & 0.2                   & 0.0                   & 6.8                   & 53.2                  & 54.7                  \\ \hline
0024$-$7204I                        & 0.229792 & 0.03845 & 6.3E-5 & 12.9                  & UL & 1218                  & 25                                  & 2.8                   & 0.0                   & 0.0                   & 2.9                   & 125.6                 & 127.0                 \\ \hline
1957+2516                         & 0.238145 & 0.28335 & 2.8E-5 & 96.6                  & ?? & 1272                  & 17                                  & 2.8                   & 0.0                   & 0.0                   & 2.8                   & 129.5                 & 129.5                 \\ \hline
\cellcolor[HTML]{C0C0C0}0751+1807 & 0.263144 & 0.39662 & 3.3E-6 & 128.3                 & He & 1370                  & 16                                  & 2.4                   & 0.0                   & 0.0                   & 2.4                   & 150.8                 & 150.8                 \\ \hline
1446$-$4701                         & 0.277666 & 0.06401 & 2.1E-5 & 19.0                  & UL & 1384                  & 23                                  & 2.1                   & 0.0                   & 0.0                   & 2.1                   & 173.0                 & 173.6                 \\ \hline
\cellcolor[HTML]{C0C0C0}0610$-$2100 & 0.286016 & 0.07349 & 3.0E-5 & 21.4                  & UL & 1412                  & 23                                  & 2.0                   & 0.0                   & 0.0                   & 2.0                   & 181.7                 & 182.1                 \\ \hline
1731$-$1847                         & 0.311134 & 0.12016 & 2.9E-5 & 33.3                  & UL & 1498                  & 22                                  & 1.7                   & 0.0                   & 0.0                   & 1.7                   & 208.2                 & 208.4                 \\ \hline
1952+2630                         & 0.391879 & 2.79820 & 4.1E-5 & 925.7                 & CO & 2062                  & 3                                   & 1.6                   & 0.0                   & 0.0                   & 1.6                   & 219.6                 & 219.6                 \\ \hline
1816+4510                         & 0.360893 & 0.59541 & 7.8E-6 & 158.2                 & He & 1702                  & 15                                  & 1.4                   & 0.0                   & 0.0                   & 1.4                   & 251.9                 & 251.9                 \\ \hline
\cellcolor[HTML]{C0C0C0}1738+0333 & 0.354791 & 0.34343 & 3.4E-7 & 89.5                  & He & 1657                  & 18                                  & 1.4                   & 0.0                   & 0.0                   & 1.4                   & 252.5                 & 252.6                 \\ \hline
1757$-$5322                         & 0.453311 & 2.08653 & 4.0E-6 & 556.7                 & CO & 2142                  & 3                                   & 1.1                   & 0.0                   & 0.0                   & 1.1                   & 315.0                 & 315.0                 \\ \hline
0024$-$7204U                        & 0.429106 & 0.52695 & 1.5E-4 & 122.8                 & He & 1895                  & 16                                  & 1.1                   & 0.0                   & 0.0                   & 1.1                   & 341.5                 & 341.5                 \\ \hline
2214+3000                         & 0.416633 & 0.05908 & 8.2E-6 & 13.3                  & UL & 1811                  & 24                                  & 1.1                   & 0.0                   & 0.0                   & 1.1                   & 341.6                 & 342.3                 \\ \hline
1431$-$4715                         & 0.449739 & 0.55006 & 2.3E-5 & 124.3                 & He & 1956                  & 16                                  & 1.0                   & 0.0                   & 0.0                   & 1.0                   & 369.1                 & 369.1                 \\ \hline
\end{tabular}
\caption{
List of potential rotation-powered binary pulsars with tiny orbital eccentricities where the
{\ttfamily ELL1k} timing model may become relevant.
This list shows that there are \emph{six} systems where $\tau_{\dot \omega_{\text {GR}} }$ is less than 
50 years. There are few PTA pulsars in the list and clearly 
{\texttt{ELL1k}}, 
at present, may not be 
relevant for their timing (PTA pulsars are marked by the  {\colorbox[HTML]{C0C0C0}{gray background}} ).
We use the following notations to denote various companions:
UL=ultra-low mass, He=He white dwarf, CO=C-O white dwarf, MS=Main sequence, ??=Unknown. 
The companion radius arises from the Roche lobe radius arguments only for the first system 
and its $R_c$ value is marked with the  {\bf boldface} font. 
The star ($\star$) symbol marks the 
cases where the main sequence mass-radius relation is used to estimate $R_c$ value.
}
\label{tab:omdot_pulsars}
\end{table*}


\subsection{On the effects of Red Timing Noise for our {\tempo} simulations }
\label{sec:RedNoise}
In this section, we investigate how the presence of red timing noise (RN) affects the measurability of $\dot\omega$.
The red timing noise
 refers to unexplained slow, stochastic wandering of TOAs observed in many pulsars \citep{Arzoumanian1994}. 
These modulations have a power-law spectrum with a low-frequency cutoff and their power spectral density takes the form \citep{Lasky2015}
\begin{equation}
\Phi_{\text{RN}}(f)=A\left(1+\frac{f^{2}}{f_{c}^{2}}\right)^{-{\rho}/{2}} \,,
\end{equation}
where $A$ is the spectral density amplitude, $f_c$ is the low-frequency cutoff and $\rho$ is the power-law index. 
The spectral power associated with the above spectral density is
\begin{equation}
P_{\text{RN}} = \int_0^\infty \Phi_{\text{RN}}(f) \, df 
 = Af_{c}\frac{\sqrt{\pi}\Gamma\left(\frac{\rho-1}{2}\right)}{2\Gamma\left(\frac{\rho}{2}\right)}
\end{equation}
when $\rho>1$.

\begin{figure}
\centering	
\includegraphics[scale=0.7]{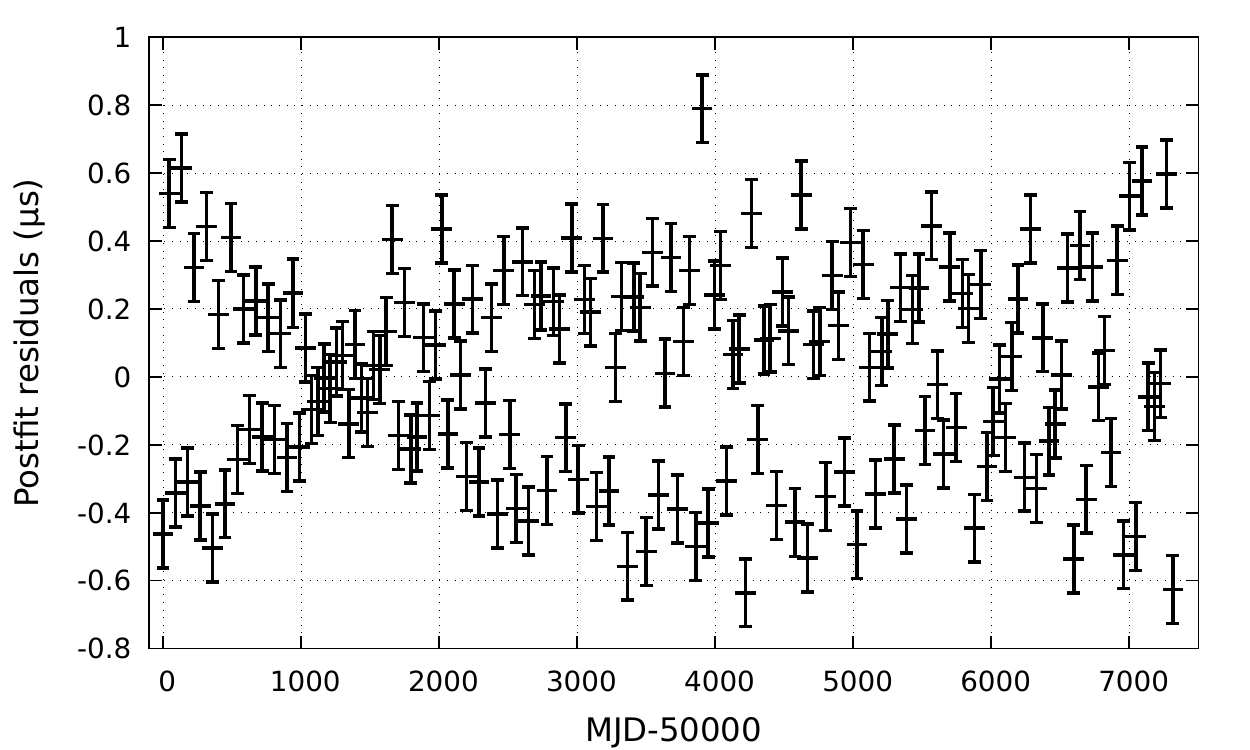}
	\caption{Post-fit timing residuals after applying harmonic whitening to the simulated data responsible for 
    Figure \ref{fig:J1919-1438_20yr-after_fdot} (Reduced $\chi^2=8.89$)}
	\label{fig:J1919-1438_2yr-HW}
\end{figure}

 The long-timescale temporal variations in timing residuals, visible in Figure \ref{fig:J1919-1438_20yr-after_fdot},
 can be mimicked by  TOA measurements affected by red timing noise.
 In Figure \ref{fig:J1919-1438_2yr-HW}, we demonstrate the consequence of red timing noise while probing the implications of 
 {our {\ttfamily ELL1k} timing model}. 
To obtain  Figure \ref{fig:J1919-1438_2yr-HW}, we assume the presence of red timing noise in the simulation data that provided 
Figure \ref{fig:J1919-1438_20yr-after_fdot}. Thereafter, we  apply harmonic whitening, detailed in \cite{Hobbs2004},
to remove it. This ensures that the long-timescale variation due to the `extra term' in equation (\ref{eq:Romer_Delay_ELL1}) will also be removed 
as evident from Figure \ref{fig:J1919-1438_2yr-HW}.
Interestingly, there are traces of  systematic variations present in the residuals of  Figure \ref{fig:J1919-1438_2yr-HW}. 
These variations in timing residuals may be associated with 
$\epsilon_2 \sin(2\Phi)$ and 
$\epsilon_1 \cos(2\Phi)$ 
terms in equation (\ref{eq:Romer_Delay_ELL1}) which vary in the orbital timescale while  the overall envelope reflects the periastron advance timescale.
These inferences force us to speculate a possible origin, in principle, for the observed red timing noise present in 
tiny-$e$ binary pulsars. 
It will be interesting to explore what fraction of the red noise arises from the  unaccounted effect of 
$\dot\omega$  while determining their ephemeris. 



 We move on to quantify how accurately we can measure  $\dot\omega$ in the presence of red timing noise.
 For this purpose, we injected several instances of red noise with different spectral parameters 
 with the help of the {\ttfamily simRedNoise} plug-in of {\tempo} into TOAs simulated using the {\ttfamily fake} plug-in. 
The simulations, as expected, were done for a PSR J1719$-$1438 like system while incorporating the effect of $\dot\omega$ using
the {\ttfamily ELL1k} timing model.
Invoking the  {\ttfamily FITWAVES} routine of {\ttfamily TEMPO2} \citep{Hobbs2004}, we applied the harmonic whitening to the 
simulated TOAs in order to remove the red noise.
Thereafter, we 
tried to recover the $\dot\omega$ value by fitting the whitened data with {the} {\ttfamily ELL1k} timing model.
We also tried to fit the whitened data with the {\ttfamily ELL1} model to see if the systematic timing residuals due to 
{our modifications} are washed out by the red noise.
The results of such {\tempo} experiments are summarized in Table \ref{tab:RedNoise} 
where we also list the parameters of the injected red noise.

\begin{table*}
\begin{tabular}{|c|c|c|c|c|c|c|c|c|c|c|c|c|}
\hline 
\multicolumn{3}{|c|}{Injected RN parameters} & \multicolumn{10}{c|}{Fitting Results}\tabularnewline
\hline 
\multirow{2}{*}{$P_{\text{RN}}$} & \multirow{2}{*}{$f_{c}$} & \multirow{2}{*}{$\rho$} & \texttt{ELL1} & \texttt{ELL1}+\texttt{FW} & \multicolumn{4}{c|}{\texttt{ELL1k}} & \multicolumn{4}{c|}{\texttt{ELL1k}+\texttt{FW}}\tabularnewline
\cline{4-13} 
 &  &  & $\chi^{2}$/dof & $\chi^{2}$/dof & $\chi^{2}$/dof & $\dot{\omega}_{\text{fit}}$ & $\sigma_{\dot{\omega}_{\text{fit}}}$ & $\varepsilon_{\dot{\omega}}$ & $\chi^{2}$/dof & $\dot{\omega}_{\text{fit}}$ & $\sigma_{\dot{\omega}_{\text{fit}}}$ & $\varepsilon_{\dot{\omega}}$\tabularnewline
(s$^{2}$ yr$^{-2}$) & (yr$^{-1}$) &  &  &  &  & ($^{\circ}$/yr) & ($^{\circ}$/yr) & (\%) &  & ($^{\circ}$/yr) & ($^{\circ}$/yr) & (\%)\tabularnewline
\hline 
\hline 
0  & - & - & 67.4  & 15.5  & 0.9  & 13.41  & 0.06  & 0.82  & 0.9  & 13.39  & 0.07  & 0.67 \tabularnewline
\hline 
\hline 
1E-29  & 0.100  & 3.5  & 68.4  & 16.8  & 1.1  & 13.43  & 0.07  & 0.97  & 0.9  & 13.48  & 0.07  & 1.32 \tabularnewline
\hline 
1E-29  & 0.037  & 3.5  & 60.9  & 16.1  & 1.1  & 13.17  & 0.07  & 0.95  & 1.0  & 13.27  & 0.08  & 0.25 \tabularnewline
\hline 
1E-29  & 0.010  & 3.5  & 67.1  & 16.0  & 1.1  & 13.18  & 0.06  & 0.88  & 1.1  & 13.15  & 0.08  & 1.12 \tabularnewline
\hline 
\hline 
1E-28  & 0.100  & 3.5  & 59.3  & 16.9  & 4.8  & 13.12  & 0.14  & 1.35  & 1.1  & 13.33  & 0.08  & 0.23 \tabularnewline
\hline 
1E-28  & 0.037  & 3.5  & 86.1  & 16.7  & 1.7  & 13.37  & 0.08  & 0.54  & 1.0  & 13.33  & 0.07  & 0.21 \tabularnewline
\hline 
1E-28  & 0.010  & 3.5  & 73.5  & 16.4  & 1.2  & 13.34  & 0.07  & 0.31  & 1.2  & 13.24  & 0.08  & 0.47 \tabularnewline
\hline 
\hline 
1E-27  & 0.100  & 3.5  & 72.9  & 16.3  & 31.6  & 13.31  & 0.45  & 0.09  & 1.8  & 13.35  & 0.11  & 0.39 \tabularnewline
\hline 
1E-27  & 0.037  & 3.5  & 121.2  & 16.2  & 9.0  & 14.30  & 0.17  & 7.55  & 1.2  & 13.39  & 0.08  & 0.69 \tabularnewline
\hline 
1E-27  & 0.010  & 3.5  & 81.8  & 16.0  & 2.1  & 13.14  & 0.09  & 1.23  & 1.0  & 13.19  & 0.08  & 0.79 \tabularnewline
\hline 
\hline 
1E-26  & 0.100  & 3.5  & 748.6  & 22.6  & 224.8  & 20.74  & 0.44  & 55.90  & 8.0  & 13.38  & 0.22  & 0.62 \tabularnewline
\hline 
1E-26  & 0.037  & 3.5  & 111.7  & 18.2  & 85.9  & 13.94  & 0.81  & 4.84  & 2.0  & 13.30  & 0.10  & 0.03 \tabularnewline
\hline 
1E-26  & 0.010  & 3.5  & 87.3  & 16.8  & 7.0  & 13.27  & 0.19  & 0.22  & 1.0  & 13.39  & 0.08  & 0.69 \tabularnewline
\hline 
\hline 
1E-25  & 0.100  & 3.5  & 4043  & 67.9  & 3311  & 18.05  & 1.89  & 35.69  & 52.8  & 13.16  & 0.55  & 1.02 \tabularnewline
\hline 
1E-25  & 0.037  & 3.5  & 1312  & 25.2  & 564.4  & 17.52  & 0.75  & 31.71  & 8.2  & 13.21  & 0.21  & 0.66 \tabularnewline
\hline 
1E-25  & 0.010  & 3.5  & 122.1  & 17.2  & 48.7  & 18.28  & 0.68  & 37.43  & 1.2  & 13.28  & 0.08  & 0.16 \tabularnewline
\hline 
\hline
1E-24 & 0.100 & 3.5 & 45449 & 863 & 23828 & 19.66 & 1.00 & 47.83 & 845 & 13.50 & 2.37 & 1.50 \\ \hline
1E-24 & 0.037 & 3.5 & 12942 & 57.1 & 5979 & 19.19 & 0.87 & 44.28 & 41.9 & 13.21 & 0.49 & 0.66 \\ \hline
1E-24 & 0.010 & 3.5 & 1201 & 18.2 & 568 & 17.05 & 0.93 & 28.17 & 2.9 & 13.26 & 0.13 & 0.30 \\ \hline
 \hline
1E-23 & 0.100 & 3.5 & 635501 & 9269 & 362386 & 24.40 & 0.81 & 83.49 & 9221 & 10.30 & 12.77 & 22.52 \\ \hline
1E-23 & 0.037 & 3.5 & 193171 & 553 & 66253 & 24.48 & 0.47 & 84.08 & 529 & 13.82 & 1.47 & 3.94 \\ \hline
1E-23 & 0.010 & 3.5 & 38419 & 50.1 & 8612 & 19.42 & 0.45 & 46.00 & 34.1 & 13.48 & 0.44 & 1.37 \\ \hline
\end{tabular}

\caption{Results which probe the effect of red noise (RN) for a PSR J1719$-$1438 like
system. 
The simulations were obtained by injecting several instances of red noise into TOAs generated using a white noise amplitude of 100 ns and $\dot{\omega}$ value
of 13.3$^{\circ}$/yr. Total simulated duration is 28 
years with one observation 
in every 45 days. 
The listed results were obtained
by fitting the simulated TOAs with different configurations -- \texttt{ELL1}
without whitening, \texttt{ELL1} after harmonic whitening, \texttt{ELL1k}
without whitening and \texttt{ELL1k} after harmonic whitening using 15 harmonics.
\texttt{FW} stands for harmonic whitening using \texttt{FITWAVES}. 
For fits
using \texttt{ELL1k}, the measured $\dot{\omega}$ values ($\dot{\omega}_{\text{fit}}$)
and the $1\sigma$ uncertainty reported by \texttt{TEMPO2} ($\sigma_{\dot{\omega}_{\text{fit}}}$)
are also listed. $\varepsilon_{\dot{\omega}}$ is the percentage error
in $\dot{\omega}_{\text{fit}}$ compared to the true value of $\dot{\omega}$
computed as $\left|\frac{\dot{\omega}_{\text{fit}}-\dot{\omega}_{\text{true}}}{\dot{\omega}_{\text{true}}}\right|\times100\%$.
\protect \\
Values quoted for \texttt{ELL1} and \texttt{ELL1}+\texttt{FW} were obtained
after fitting for $f$, $\dot{f}$, $\epsilon_{10}$, $\epsilon_{20}$,
$\dot{\epsilon}_{1}$ and $\dot{\epsilon}_{2}$. Values quoted for
\texttt{ELL1k} and \texttt{ELL1k}+\texttt{FW} were obtained after fitting
for $f$, $\dot{f}$, $\epsilon_{10}$, $\epsilon_{20}$, and $\dot{\omega}$
with an initial guess value of $12^{\circ}$/yr for $\dot{\omega}$. 
A careful inspection of various columns of reduced $\chi^{2}$ ($\chi^{2}$/dof) reveals the usefulness of the {\ttfamily ELL1k} model even in the presence of substantial red noise.
}
\label{tab:RedNoise}
\end{table*}

\begin{figure}
\includegraphics[scale=0.7]{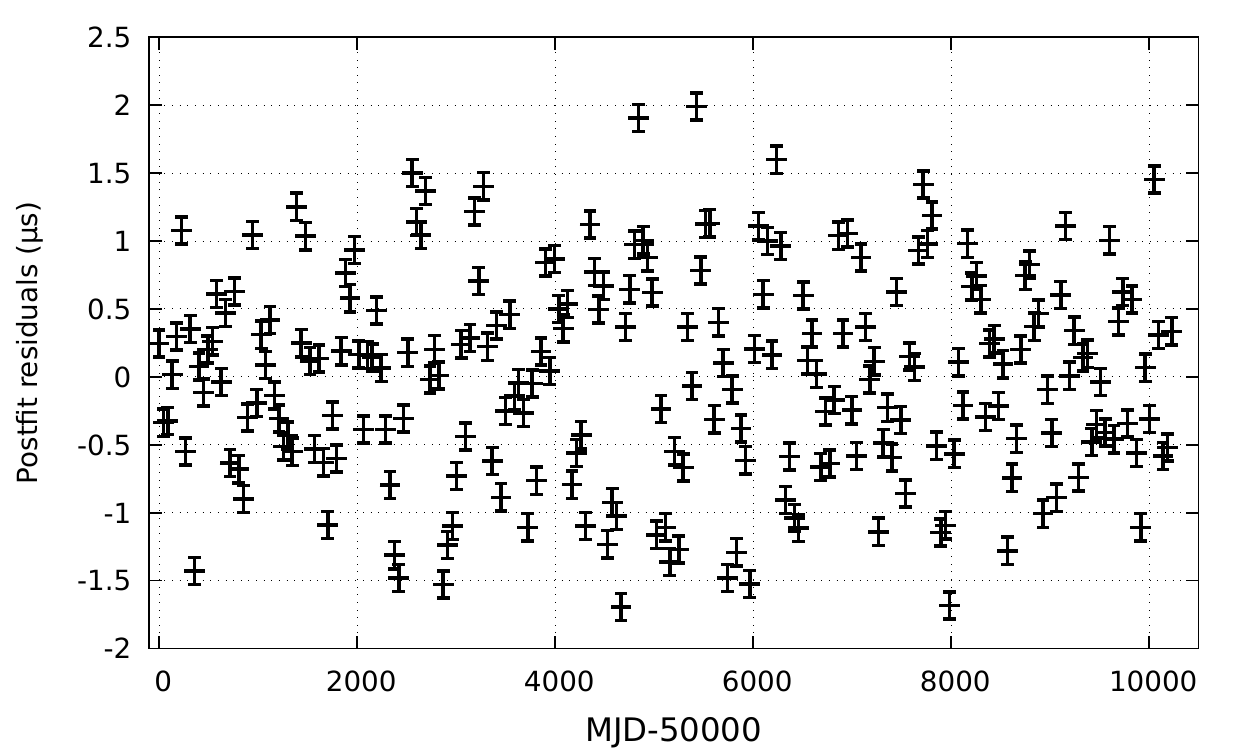}
\caption{Post-fit timing residuals after applying harmonic whitening to simulated TOAs injected with red noise having spectral parameters $P_{\text{RN}}=10^{-25}$ s$^2$ yr$^{-2}$, $f_c=0.1$ yr$^{-1}$ and $\rho=3.5$ (Reduced $\chi^2=58.7$). 
Total duration of our {\tempo} simulations is 28 years with 
\good{one} observation in every 45 days. 
Note the absence of systematic timing residuals such as those seen in Figure \ref{fig:J1919-1438_2yr-HW}.}
\label{fig:J1919-1438_TooMuchRN}
\end{figure}

From Table \ref{tab:RedNoise} it is clear that {\ttfamily ELL1k} gives better reduced $\chi^2$ values than {\ttfamily ELL1} in all cases, 
indicating that red noise does not completely wash out the effects introduced by {our modifications}. 
However, when the red noise power is large ($P_{\text{RN}}\gtrsim 10^{-25}$ in Table \ref{tab:RedNoise}), 
the systematic timing residuals such as those seen in Figure \ref{fig:J1919-1438_2yr-HW} are no longer visually identifiable post harmonic whitening (see Figure \ref{fig:J1919-1438_TooMuchRN}). 
Therefore, the decision to employ 
the {\ttfamily ELL1k} model over {\ttfamily ELL1} model should be made after estimating $\tau_{\dot\omega}$ values rather than by visual inspection of timing residuals.

It can be seen from Table \ref{tab:RedNoise} that the value of $\dot\omega$ can be 
\good{obtained} with reasonable ($\sim 1\%$) accuracy by applying harmonic whitening when the red noise amplitude is sufficiently small. It is still possible to obtain an estimate for $\dot\omega$  even  for higher red noise amplitudes albeit with  poorer accuracies and higher reduced $\chi^2$ values. 
Answer to the question how small an amplitude is ``sufficiently small'' crucially depends on $f_c$ as the accuracy of $\dot\omega$ measurement  reduces significantly for larger values of $f_c$. 
This is to be expected as harmonic whitening removes low-frequency noise more effectively than high-frequency noise. 

A perhaps surprising observation is that applying harmonic whitening does not affect the accuracy of the $\dot\omega$ measurement significantly. 
This is due to the fact that the timing residuals introduced by 
the exact secular evolution of LL parameters vary in two distinct timescales \good{(i.e., orbital and periastron advance)} as we have noticed earlier in this section. 
Although the long-timescale variation introduced by the `extra term' is removed by \texttt{FITWAVES}, the contributions
which arise from the general temporal evolutions of the LL parameters should 
allow us to obtain a $\dot\omega$ estimate even in the presence of red noise. 
This point is further emphasized by the observation that the accuracy of $\dot\omega$ measurement does not decrease even when the red noise frequency cutoff corresponds to the periastron advance timescale itself ($f_c=0.037$ yr$^{-1} \sim \tau_{\dot\omega}^{-1}$).

\section{Summary and Discussions}
\label{sec:discussion}
 We explored the possibility of measuring the effect of periastron advance in small-eccentricity  binary pulsars. 
This was pursued by implementing a timing model, {\ttfamily ELL1k}, in {\tempo} that essentially incorporated 
general temporal evolutions for the crucial LL parameters present in the {\ttfamily ELL1} timing model.
This present {prescription} should be relevant for binary pulsars that can, in principle, experience significant periastron advance.
 With the help of {\tempo} simulations and using PSR J1719$-$1438 as an example, we probed the observational implications of {the} {\ttfamily ELL1k} timing model
 in comparison with the {\tempo} implementation of Wex's {\ttfamily ELL1} model.
 This comparison provides significant timing residuals while simulating a decade-long timing of PSR J1719$-$1438 even while using the conservative 
 $\dot{\omega}_{\text{GR}}$ estimate for the periastron advance.
 Our {\tempo} simulations suggest the possibility of measuring the effect of $\dot\omega$ in PSR J1719$-$1438 by pursuing a high-cadence decade-long timing campaign.
 Additionally,  we investigated  the measurability
of periastron advance in the presence of red timing noise in our {\tempo} simulations for PSR J1719$-$1438.
This is achieved 
by injecting several instances of red noise with varying spectral properties into the simulated TOAs. 
We concluded that $\dot\omega$ can be measured to about 1\% accuracy provided the amplitude of the red noise is sufficiently small. 
Furthermore, the tolerable amplitude of red-noise increases with the red-noise timescale ($1/f_c$).
 Interestingly, if the measured $\dot\omega$ turns out to be larger than the expected $\dot\omega_\text{GR} \sim$ 13$\degree$/yr, 
 it should lead to constraints on the apsidal motion constant $k_2$ of its unique companion. 
We also compiled a list of binary pulsars with tiny orbital
eccentricities 
where $\dot\omega$ effects may become noticeable  within a reasonable timescale especially if the classical contributions to 
periastron advance dominates over their GR counterpart.
More importantly, we showed that the linear-in-time evolutions of LL parameters do not introduce 
any significant timing residuals in the currently employed PTA pulsars with tiny orbital eccentricities. 

 The present effort should be interesting for the following scenarios.
 The first one involves the detection of sub-$\mu$Hz GWs using an ensemble of MSPs in a PTA experiment 
 that demands post-fit timing residuals of the order of 10 nano-seconds. 
 Unfortunately, this is not attainable for most of the current PTA MSPs and many of them are part of nearly circular binary systems.
  A possible source of higher post-fit timing residuals can be unmodeled systematic effects as evident from our plots.
  Therefore,  the use of a more refined timing model can, in principle, lower post-fit residuals,
especially for MSPs in compact orbits with small eccentricities.
However,  it is unlikely that {the} {\ttfamily ELL1k} model will lead to any improvements to the current PTA sensitivity and
this is mainly due to the large $\tau_{{\dot \omega}}$ values for the current list of PTA pulsars.
However, it will be helpful to check the relevance of {this} timing model while including new MSPs into the existing list of PTA pulsars.
The second scenario involves accreting millisecond X-ray pulsars like SAX J1808.4$-$3658.
It should be interesting to explore the implications of {this} model while analyzing timing data associated with its many observed outbursts.
For such accretion-powered systems, the classical contributions to $\dot \omega$ can be quite large and therefore it may be worthwhile to
perform coherent timing of its outburst data by employing our timing model. 
This may, in principle,  lead to an estimate for the $k_2$ value of its brown-dwarf companion and detailed 
analysis and simulations will be required to quantify our suggestion.
Note that it will require LISA observations of eccentric galactic binaries for estimating $k_2$ values of 
degenerate objects \citep{Willems2008}.
Finally, {this} timing model should be relevant during the FAST-SKA era as  
the MSP population is expected to quadruple during this era \citep{Levin2017}.
It is reasonable to expect that SKA will monitor dozens of short orbital period MSPs 
in nearly circular orbits and 
{the} {\ttfamily ELL1k} timing model will be required for the high-cadence timing of such systems.

\section*{Acknowledgements}
We are grateful to Norbert Wex for his valuable comments and suggestions and
thank the anonymous referee for her/his constructive comments
which helped us to improve the manuscript. 
Additionally, we would like to express our gratitude to George Hobbs and Michael Keith 
for promptly incorporating our {\ttfamily ELL1k} timing model into the {\tempo} 
pulsar timing package.




\appendix

\section{Expressions required to implement \texttt{ELL1\lowercase{k}} model in \tempo}
\label{sec:t2_inputs}

We list below the partial derivatives of the \Romer delay, namely  equation (\ref{eq:Romer_Delay_ELL1}), with respect to the relevant binary
 parameters while  neglecting terms of $\order(e)$ and $\order({\dot P}_b)$.
These expressions are 
 required  to implement {the \texttt{ELL1k}} timing model in {\tempo} and in Table \ref{tab:ELL1K_params} we provide a 
 \good{list of} the binary parameters and their {\tempo} notations.

\begin{subequations}
\begin{align}
\frac{\partial\Delta_R}{\partial P_b} &= -\frac{x}{c}n_{b}\frac{\tau}{P_{b}}\cos\Phi \,, \\
\frac{\partial\Delta_R}{\partial \dot{P}_b} &= -\frac{1}{2}\frac{x}{c}n_{b}\frac{\tau^{2}}{P_{b}}\cos\Phi  \,,\\
\frac{\partial\Delta_R}{\partial (x/c)} &= \sin \Phi \,,\\
\frac{\partial\Delta_R}{\partial (\dot{x}/c)} &= \tau\sin \Phi \,,\\
\frac{\partial\Delta_R}{\partial \epsilon_{10}} &=  -\frac{x}{2c}\left(\left(\cos2\Phi+3\right)\cos\dot{\omega}\tau+\sin2\Phi\;\sin\dot{\omega}\tau\right)  \,,\\
\frac{\partial\Delta_R}{\partial \epsilon_{20}} &=  -\frac{x}{2c}\left(\left(\cos2\Phi+3\right)\sin\dot{\omega}\tau-\sin2\Phi\;\cos\dot{\omega}\tau\right)  \,,\\
\frac{\partial\Delta_R}{\partial {\dot{\omega}}} &= -\frac{x}{2c}\left(\epsilon_{2}\left(\cos2\Phi+3\right)+\epsilon_{1}\sin2\Phi\right)\tau   \,,\\
\frac{\partial\Delta_R}{\partial T_{\ascnode}} &= -\frac{x}{c} n_{b} \cos \Phi  \,.
\end{align}
In addition, if one wishes to fit for the logarithmic derivative of eccentricity $\xi$, the corresponding partial derivative is given by
\begin{align}
\frac{\partial\Delta_R}{\partial \xi} &=  \frac{x}{2c}\left(\epsilon_{2}\sin2\Phi-\epsilon_{1}\left(\cos2\Phi+3\right)\right)\tau \,.
\end{align}
\end{subequations}
Note that $\xi$ is not present in the standard list of {\tempo} parameters.

\begin{table}
\begin{tabular}{|c|c|c|}
\hline 
Parameter &  {\tempo} parameter name & Unit\tabularnewline
\hline 
\hline 
$P_{b}$ & PB & days\tabularnewline
\hline 
$\dot{P}_{b}$ &  PBDOT & \tabularnewline
\hline 
$x$ &  A1 & lt-s\tabularnewline
\hline 
$\dot{x}$ &  A1DOT & lt-s/s\tabularnewline
\hline 
$\epsilon_{10}$ &  EPS1 & \tabularnewline
\hline 
$\epsilon_{20}$ &  EPS2 & \tabularnewline
\hline 
$\dot{\omega}$ &  OMDOT & $\degree$/yr\tabularnewline
\hline 
$\xi$ &   & yr$^{-1}$ \tabularnewline
\hline 
$T_{\ascnode}$ & TASC & MJD\tabularnewline
\hline 
\end{tabular}
\caption{Binary model parameters for {\ttfamily ELL1k}}
\label{tab:ELL1K_params}
\end{table}

\section{\good{Comparison of Timing Models for Low-Eccentricity Pulsar Binaries}}
\label{sec:ell1_comparison_table}
\good{
With our proposed \texttt{ELL1k} model described in this work, there are now four different timing models for low-eccentricity binaries.  We provide in Table \ref{tab:timing_model_comparison} a brief comparison of these models.
}

\begin{table*}
\begin{tabular}{|m{2cm}||m{3.5cm}||m{11cm}|}
\hline 
Binary Model & Model Parameters & Description\tabularnewline
\hline 
\hline 
\texttt{ELL1} & $P_{b}$, $\dot{P}_{b}$, $x$, $\dot{x}$, $\epsilon_{10}$, $\epsilon_{20}$,
$\dot{\epsilon}_{1}$, $\dot{\epsilon}_{2}$, $T_{\ascnode}$, $m_{c}$,
$\sin\iota$ & Model for low-eccentricity binaries. Uses Laplace-Lagrange parameters
and $T_{\ascnode}$ instead of usual Keplerian parameters $e$, $\omega$
and $T_{0}$. Incorporates $\mathcal{O}(e)$ corrections of  the \Romer
and Shapiro delays.\newline
Reference : \citet{Lange2001} \tabularnewline 
\hline 
\texttt{ELL1H} & $P_{b}$, $\dot{P}_{b}$, $x$, $\dot{x}$, $\epsilon_{10}$, $\epsilon_{20}$,
$\dot{\epsilon}_{1}$, $\dot{\epsilon}_{2}$, $T_{\ascnode}$, $h_{3}$,
$h_{4}$, $\varsigma$, $N_{\text{harm}}$ & Model for low-eccentricity, low to moderate inclination binaries with measurable
Shapiro delay. Same as \texttt{ELL1} except that this model uses a
truncated Fourier expansion of Shapiro delay instead of the exact
$\mathcal{O}(e)$ expression, where $h_i$ is the amplitude of the $i$th harmonic, $\varsigma=\frac{\sin\iota}{1+|\cos\iota|}$ and $N_\text{harm}$ is the number of harmonics in the Fourier expansion.\newline
Reference : \citet{Freire2010}\tabularnewline
\hline 
\texttt{ELL1+} & $P_{b}$, $\dot{P}_{b}$, $x$, $\dot{x}$, $\epsilon_{10}$, $\epsilon_{20}$,
$\dot{\epsilon}_{1}$, $\dot{\epsilon}_{2}$, $T_{\ascnode}$, $m_{c}$,
$\sin\iota$ & Model for low-eccentricity, wide-orbit binaries where $xe^{2}/c$
is comparable to or larger than the timing precision. Incorporates
$\mathcal{O}(e^{2})$ terms in the \Romer delay.\newline
Reference : \citet{Zhu2018}\tabularnewline
\hline 
\texttt{ELL1k} & $P_{b}$, $\dot{P}_{b}$, $x$, $\dot{x}$, $\epsilon_{10}$, $\epsilon_{20}$,
$\dot{\omega}$, $T_{\ascnode}$, $m_{c}$, $\sin\iota$ & Model for low-eccentricity, compact binaries showing significant advance
of periastron. Incorporates an extra, slowly varying term in the \Romer
delay that was neglected in \texttt{ELL1}.
\newline
Reference : This work
\tabularnewline
\hline 
\end{tabular}
\caption{Comparison of different timing models for low-eccentricity binary pulsars. 
Listed references provide further details on the various models and their parameters.}
\label{tab:timing_model_comparison}
\end{table*}



\bibliographystyle{mnras}
\bibliography{ELL1}


\bsp	
\label{lastpage}

\end{document}